\documentclass[a4paper,1p,number]{elsarticle}
\usepackage{amsmath,amsfonts,amssymb,latexsym}
\usepackage{graphicx,color}

\newcommand{\eg}{{\it e.g.}}
\newcommand{\etal}{{\it et al.}}
\newcommand{\ie}{{\it i.e.}}
\newcommand{\be}{\begin{equation}}
\newcommand{\ee}{\end{equation}}
\newcommand{\bea}{\begin{eqnarray}}
\newcommand{\eea}{\end{eqnarray}}
\newcommand{\bef}{\begin{figure}}
\newcommand{\eef}{\end{figure}}
\newcommand{\bce}{\begin{center}}
\newcommand{\ece}{\end{center}}

\newcommand{\dd}{\text{d}}

\newcommand{\lsim}{\lesssim}
\newcommand{\gsim}{\gtrsim}

\newcommand{\TeV}{\mathrm{TeV}}
\newcommand{\GeV}{\mathrm{GeV}}
\newcommand{\MeV}{\mathrm{MeV}}
\newcommand{\fm}{\mathrm{fm}}
\DeclareMathOperator{\im}{Im}

\begin{document}

\title{Pseudo-Critical Enhancement of Thermal Photons in Relativistic
  Heavy-Ion Collisions?}

\author[fias,gu]{Hendrik van Hees}
\author[nanjing]{Min He}
\author[texas]{Ralf Rapp}

\address[fias]{Frankfurt Institute for Advanced Studies, Ruth-Moufang-Stra{\ss}e 1,
D-60438 Frankfurt, Germany}
\address[gu]{Institute for Theoretical Physics, Max-von-Laue-Stra{\ss}e 1,
D-60438 Frankfurt, Germany}
\address[nanjing]{Department of Applied Physics, Nanjing University of
  Science and Technology, Nanjing 210094, China} 
\address[texas]{Cyclotron Institute and Department of Physics\&Astronomy, Texas A\&M University, 
College Station, TX 77843-3366, U.S.A.}

\date{\today}

\begin{abstract}
  We compute the spectra and elliptic flow of thermal photons emitted in
  ultrarelativistic heavy-ion collisions (URHICs) at RHIC and LHC. The
  thermal emission rates are taken from complete leading-order rates for
  the QGP and hadronic many-body calculations including baryons and
  antibaryons, as well as meson-exchange reactions (including
  Bremsstrahlung). We first update previous thermal fireball
  calculations by implementing a lattice-QCD based equation of state and
  extend them to compare to recent LHC data. We then scrutinize the
  space-time evolution of Au-Au collisions at RHIC by employing an ideal
  hydrodynamic model constrained by bulk- and multistrange-hadron
  spectra and elliptic flow, including a non-vanishing initial flow. We
  systematically compare the evolutions of temperature, radial flow,
  azimuthal anisotropy and four-volume, and exhibit the temperature
  profile of thermal photon radiation. 
  Based on these insights, we put forward a scenario with a ``pseudo-critical 
  enhancement'' of thermal emission rates, and investigate its impact 
  on RHIC and LHC direct photon data.
\end{abstract}

\begin{keyword}
heavy-ion collisions \sep QCD phase diagram \sep direct photons
\PACS{25.75.-q  \sep 25.75.Dw  \sep 25.75.Nq}
\end{keyword}

\maketitle

\section{Introduction}
\label{intro}

The thermal emission rate of photons from strongly interacting matter
encodes several interesting properties of the radiating medium (see,
\eg,
Refs.~\cite{Alam:1999sc,Peitzmann:2002,Arleo:2004gn,Rapp:2009yu,Gale:2009gc}
for reviews). Its spectral slope reflects the temperature of the system
while its magnitude is related to the interaction strength of the charge
carriers. In ultrarelativistic heavy-ion collisions (URHICs), the size
of the interacting fireball is much smaller than the mean-free path of
photons. Thus, the latter can probe the hot and dense interior of the
medium. However, the observed photon spectra receive contributions from
all reaction stages, \ie, primordial $NN$ collisions, pre-equilibrium,
quark-gluon plasma (QGP) and hadronic phases, plus final-state decays of
short-lived resonances (these so-called ``direct'' photons exclude
decays of long-lived hadrons, \eg, $\pi$ and $\eta$). Calculations of
direct-photon spectra require good control over both the microscopic
emission rates and the space-time evolution of the medium. The latter
not only determines the local emission temperature, but also the
collective-flow field which generally imparts a net blue-shift on the
radiated photons. In addition, the azimuthal asymmetry of the thermal
photon spectra, $v_2^\gamma$, is of
interest~\cite{Chatterjee:2006,Liu:2009kta,Holopainen:2011pd,vanHees:2011vb,Dion:2011pp,Mohanty:2011fp,Shen:2013cca,Linnyk:2013wma}:
since the bulk $v_2$ requires several $\fm/c$ to build up, the observed
value for photons helps to further constrain their emission history.

Direct-photon spectra in URHICs have been extracted in
Pb-Pb($\sqrt{s}=0.017\, A \TeV$) collisions at the Super Proton
Synchrotron (SPS)~\cite{Aggarwal:2000th}, in Au-Au($\sqrt{s} =0.2\, A
\TeV$) at the Relativistic Heavy-Ion Collider
(RHIC)~\cite{Adare:2008fq}, and in Pb-Pb($\sqrt{s}=2.76\,A \TeV$) at the
Large Hadron Collider (LHC)~\cite{Wilde:2012wc}. At SPS, various
theoretical models could approximately reproduce the measured spectra by
adding thermal radiation from an equilibrated expanding fireball to a
primordial component estimated from pp
data~\cite{Srivastava:2000pv,Huovinen:2001wx,Turbide:2003si,Mohanty:2009cd,Bauchle:2010sr}.
The thermal yield prevailed over the primordial one up to transverse
momenta of $q_T \approx2$-$4\, \GeV$. However, a decomposition into
contributions from QGP and hadronic radiation, which would allow for a
better characterization of the origin of the signal, remains
ambiguous. By subtracting the primordial component from their data, the
PHENIX collaboration extracted the ``excess radiation'' and determined
its inverse-slope parameter (``effective temperature'') in Au-Au
collisions at RHIC as
$T_\mathrm{eff}=221\pm19^\mathrm{stat}\pm19^\mathrm{syst} \, \MeV$.
Accounting for the aforementioned blue-shift effect, this result
indicates that most of the radiation emanates from matter temperatures
$T<200 \, \MeV$, challenging the notion of early QGP
radiation~\cite{vanHees:2011vb}. A subsequent first measurement of the
direct-photon $v_2$ supports this finding~\cite{Adare:2011zr}: in the
regime where thermal radiation is expected to be large, $q_T \lsim 3
\,\GeV$, $v_2^\gamma(q_T)$ turns out to be comparable to that of pions,
which are only emitted at the end of the fireball evolution, \ie, at
thermal freezeout, $T_\mathrm{fo} \simeq 100\, \MeV$. The large
$v_2^\gamma$, also found at LHC~\cite{Lohner:2012ct}, thus puts rather
stringent constraints on the origin of the excess photons.

In previous work~\cite{vanHees:2011vb} we have calculated thermal photon
spectra at RHIC, differing from existing calculations in mainly two
aspects. First, a more extensive set of hadronic thermal photon rates
has been employed~\cite{Turbide:2003si}, which, in particular, includes
the contributions from baryons and antibaryons (known to be important in
the dilepton context~\cite{Rapp:2000pe,Rapp:2013nxa}). These rates
approximately match complete leading-order (LO) QGP rates around the
pseudo-critical temperature, $T_\mathrm{pc}\simeq 170
\,\MeV$~\cite{Arnold:2001ms}, thus rendering a near continuous
emissivity across the transition region. Second, a schematic medium
evolution was constructed utilizing a blast-wave type elliptic-fireball
model, quantitatively fit to spectra and $v_2$ of bulk hadrons ($\pi$,
K, p) at $T_\mathrm{fo} \simeq 100\,\MeV$ and multistrange hadrons
(e.g., $\phi$ and $\Omega^-$) at $T_\mathrm{ch}=170\,\MeV$. The
implementation of this ``sequential freezeout'' is phenomenologically
motivated~\cite{He:2010vw}, and, in particular, leads to a saturation of
the bulk-medium $v_2$ close to the transition regime, after about
4-$6\,\fm/c$ for central and semi-central Au-Au collisions at RHIC. As a
result, the direct-photon $v_2$ increased by a factor of $\sim 3$ over
existing calculations, reaching into the error bars of the PHENIX data.

In the present paper we expand on and scrutinize these findings by
extending the calculations to LHC energy and then employing a previously
constructed ideal hydrodynamic bulk-evolution~\cite{He:2011zx} to
conduct a detailed comparison to the emission characteristics of the
fireball.  Much like the latter, this hydro evolution has been
quantitatively constrained by bulk-hadron spectra and $v_2$, utilizing
the concept of sequential freezeout. Both evolutions will be based on a
lattice-QCD equation of state (EoS) for the QGP, matched to a hadron
resonance gas (HRG) with chemical freezeout. The comparisons will
encompass the time evolution of radial and elliptic flow, temperature,
four volume and photon emission profile. Motivated by these comparisons,
which identify the transition region as a key contributor to thermal
photon spectra, we conjecture an enhancement of the currently available
photon emission rates around $T_{\rm pc}$ and explore in how far this
could help to resolve the discrepancies with the data.
 
Our article is organized as follows. In Sec.~\ref{sec_bulk} we
recall basic ingredients and features of the fireball
(Sec.~\ref{ssec_fb}) and hydrodynamic (Sec.~\ref{ssec_hydro}) bulk
evolutions, including analyses of their time and temperature profiles
of collective flow and four volume (Sec.~\ref{ssec_comp}). In
Sec.~\ref{sec_gam} we investigate the spectra and elliptic flow of
thermal photons emitted from the fireball (Sec.~\ref{ssec_gam-fb}) and
hydro (Sec.~\ref{ssec_gam-hydro}), and (after adding
primordial production) compare to recent RHIC and LHC data. In
Sec.~\ref{sec_disc} we analyze the differences in the fireball and hydro
photon results in view of the insights from Sec.~\ref{sec_bulk} and
discuss possible origins of current discrepancies with data. We
conclude and outline future investigations in Sec.~\ref{sec_sum}.

\section{Bulk Evolution Models}
\label{sec_bulk}
The calculation of thermal-photon spectra in URHICs is based on the
differential emission rate per unit phase space from a strongly
interacting medium of temperature $T$ and baryon-chemical potential
$\mu_B$,
\begin{equation}
\begin{split}
\label{rate}
q_0\frac{dN_{\gamma}}{d^4xd^3q} = &-\frac{\alpha_\mathrm{EM}}{\pi^2}
\ f^B(q_0;T) \\
&\times \im \Pi_\mathrm{EM}^T(q_0=q;\mu_B,T) \ . 
\end{split}
\end{equation}
Here, the rate is written in terms of the 3-D transverse part of the
electromagnetic current correlator, $\Pi_\mathrm{EM}^T$, and the thermal
Bose distribution function, $f^\mathrm{B}$, where $q_0=q$ denote the
energy and three-momentum of the photon in the local rest frame of the
medium. This expression is leading order in the electromagnetic (EM)
coupling $\alpha_\mathrm{EM}$, required to produce the photon without
further EM interaction when traversing the fireball. Alternatively, one
may express the rate in terms of photon-production scattering matrix
elements, appropriately convoluted over the thermal distribution
functions of the incoming particles. This approach is usually more
convenient when evaluating tree-level diagrams (\eg, $t$-channel meson
exchanges) which are expected to prevail at high photon
energies~\cite{Turbide:2003si}.
 
The calculation of a thermal photon spectrum in URHICs requires to
integrate the above rate over the entire four-volume of the reaction,
$V_4=\int \dd^4x$, accounting for the local temperature and collective
expansion velocity of the emission point. In the following, we will
briefly recall how this four-volume integration is done in two different
models, \ie, a schematic blast-wave type fireball (Sec.~\ref{ssec_fb})
and ideal hydrodynamics (Sec.~\ref{ssec_hydro}); both are based on the
same equation of state and fits to the same set of bulk-hadron
observables. This will be followed by a detailed comparison of the flow,
temperature and four-volume profiles (Sec.~\ref{ssec_comp}).

\subsection{Thermal Fireball}
\label{ssec_fb}
The thermal fireball model is based on an isotropically expanding
cylinder with volume
\begin{equation}
V_\mathrm{FB}(t) = \pi a(t) b(t) (z_0 +ct),
\label{V_fb}
\end{equation}
where the elliptic transverse area is characterized by semi-major and
-minor axes, $a(t)$ and $b(t)$, respectively. Their initial values are
estimated from the nuclear overlap at given impact parameter, while the
initial longitudinal size, $z_0$, controls the formation time of the
thermal medium. Assuming a constant total entropy, $S_\mathrm{tot}$, of
the fireball at given collision centrality (fixed by the observed number
of charged particles), the time evolution of the entropy density follows
as $s(t)=S_\mathrm{tot}/V_\mathrm{FB}(t)$. Once the EoS is specified,
i.e., the temperature dependence of $s$, one can convert $s(t)$ into
$T(t)$. In our previous calculations of thermal-photon
spectra~\cite{vanHees:2011vb} we used a quasi-particle QGP EoS with a
first-order transition into a HRG and chemical freezeout at
$T_\mathrm{c}=T_\mathrm{ch}=180\, \MeV$~\cite{Rapp:2000pe}. Here, we
update the EoS with a fit to lattice-QCD data for the QGP
part~\cite{He:2011zx}, smoothly matched to a HRG at
$T_\mathrm{pc}=170\,\MeV$ and chemical freezeout at
$T_\mathrm{ch}=160\,\MeV$, at both RHIC and LHC
energies~\cite{Andronic:2005yp,Stachel:2013zma}. For $T<T_\mathrm{ch}$,
effective chemical potentials for pions, kaons, antibaryons etc., are
introduced~\cite{Rapp:2002fc} to preserve the finally observed hadron
ratios extracted from the chemical-freezeout fits, while strong
processes (\eg, $\pi\pi\leftrightarrow\rho$) are assumed to maintain
chemical equilibrium (so-called \emph{partial} chemical equilibrium).
Following Ref.~\cite{He:2011zx} we refer to this EoS as
``\textit{latPHG}''.

\begin{figure}[!t]
\begin{minipage}{0.48\linewidth}
\includegraphics[width=\textwidth]{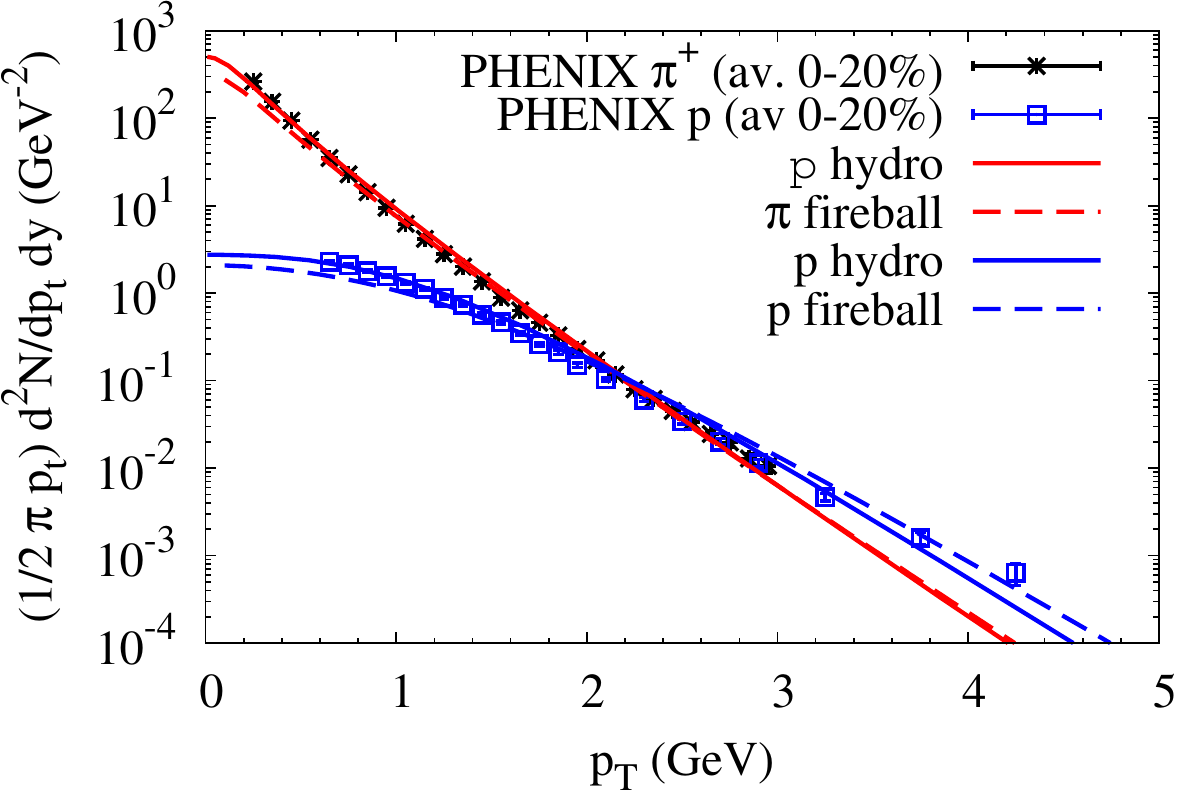}
\end{minipage}\hfill
\begin{minipage}{0.48\linewidth}
\includegraphics[width=\textwidth]{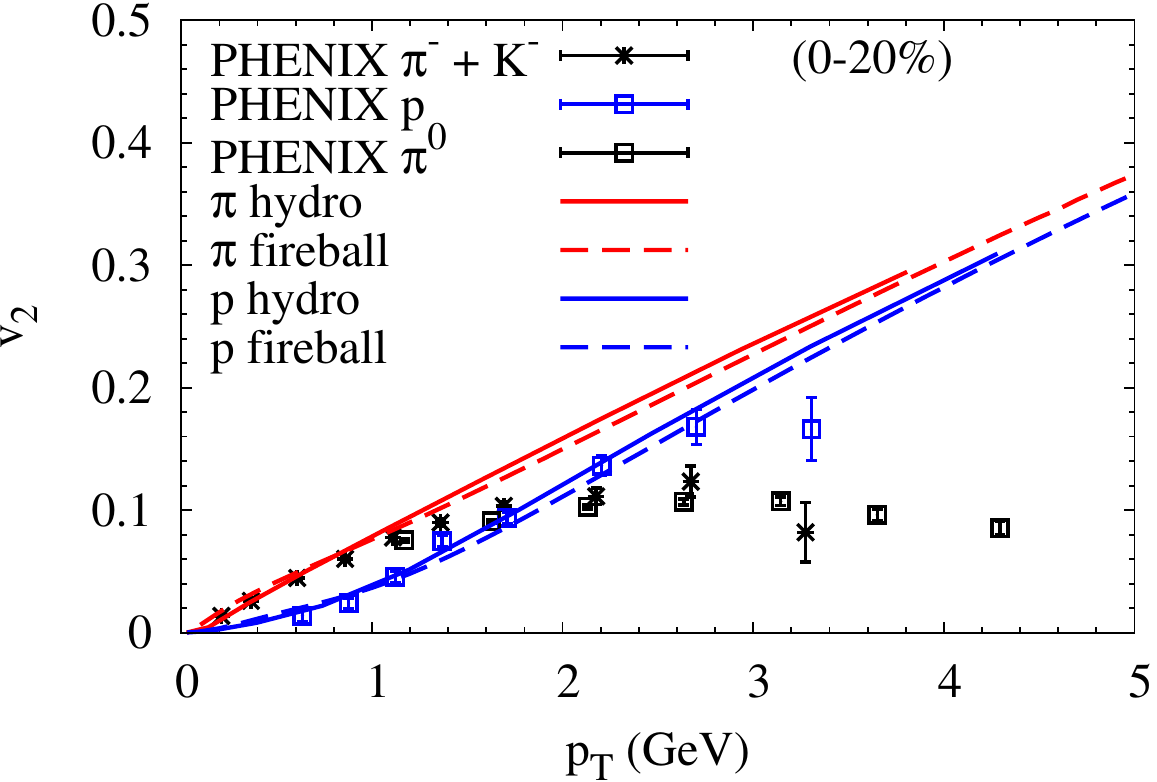}
\end{minipage}
\begin{minipage}{0.48\linewidth}
\includegraphics[width=\textwidth]{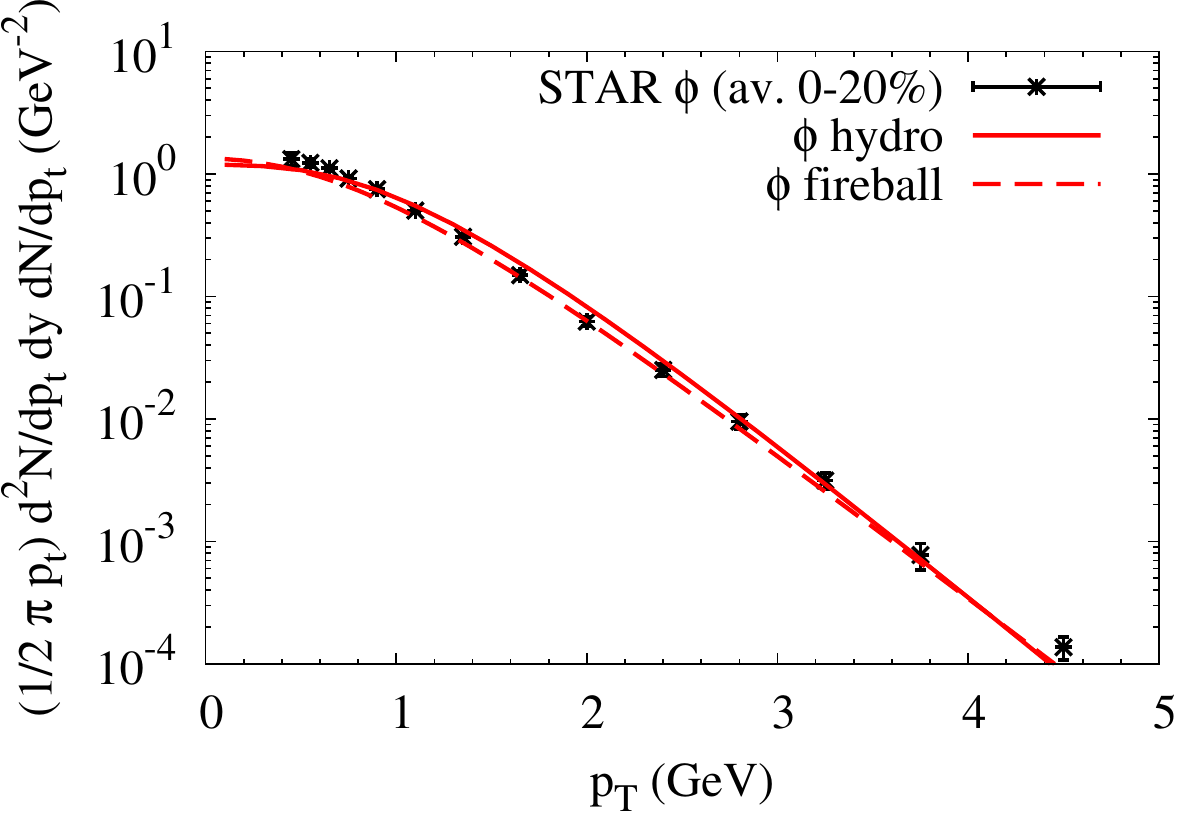}
\end{minipage}\hfill
\begin{minipage}{0.48\linewidth}
\includegraphics[width=\textwidth]{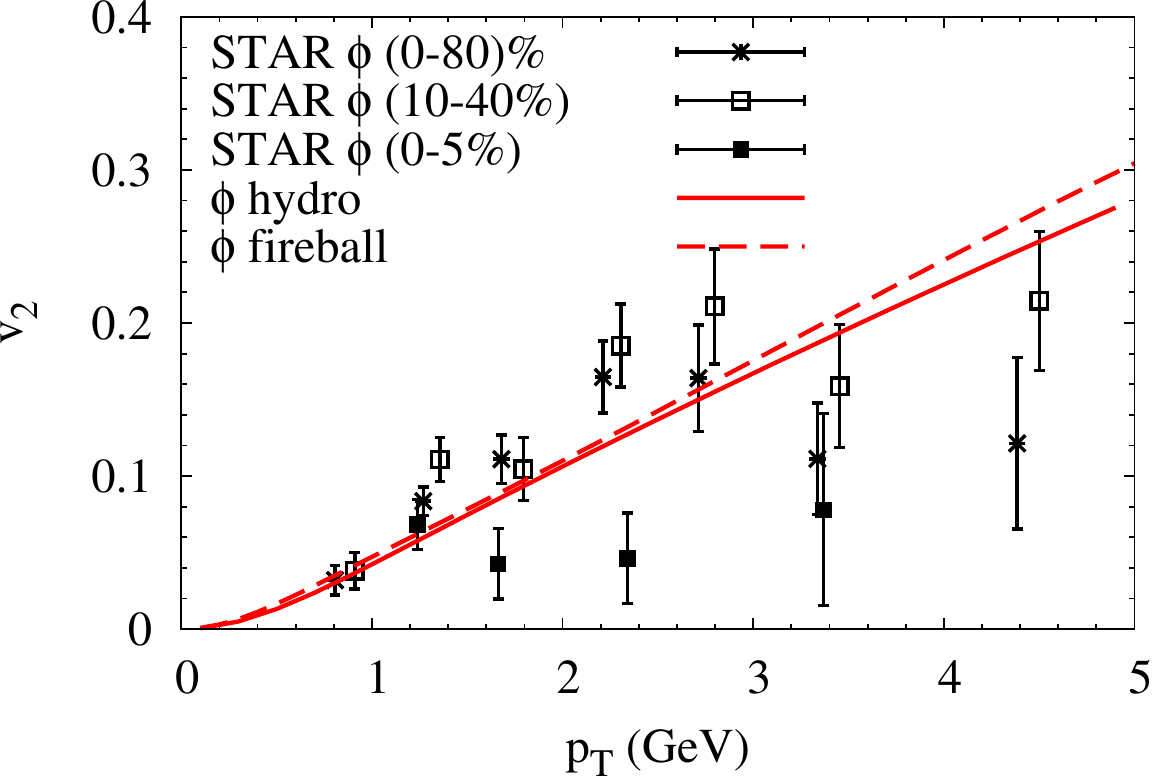}
\end{minipage}
\caption{(Color online) Fits to the spectra and elliptic flow of light
  hadrons ($T_\mathrm{fo}\simeq110\,\MeV$, upper two panels) and $\phi$
  mesons ($T_\mathrm{fo}=160\,\MeV$, lower two panels) in
  Au-Au($\sqrt{s}=200\,A \GeV$) collisions, using EoS \textit{latPHG}
  within either the fireball (dashed lines) or ideal hydrodynamic model 
  (solid lines). The data are taken from
  Refs.~\cite{Abelev:2007rw,Adler:2003kt,Adler:2003cb,Adams:2003xp}.}
\label{fig_hadrons}
\end{figure}

With this set-up, the time dependence of the elliptic radii, $a(t)$ and
$b(t)$, can be constructed with guidance from hydrodynamic
models~\cite{Kolb:2003dz} to approximately reproduce their time
evolution of the radial and elliptic flow, as well as momentum-space
anisotropy~\cite{vanHees:2011vb}. In addition, the idea of sequential
freeze-out has been implemented, \ie, a kinetic decoupling of
multistrange hadrons (\eg, $\phi$ and $\Omega^-$) at chemical
freezeout. This requires a somewhat faster transverse expansion than in
the original hydro models~\cite{Kolb:2003dz}, but is consistent with the
observed phenomenon of constituent-quark number scaling of elliptic
flow. Importantly, it implies that the bulk-$v_2$ essentially saturates
close to $T_\mathrm{c}\simeq T_\mathrm{ch}$. With a suitable choice in
initial conditions this can be recovered in hydrodynamic
simulations~\cite{He:2011zx}, as we will see below. For a more accurate
reproduction of the final-state hadron multiplicities compared to our
previous work~\cite{vanHees:2011vb} (accounting for the modified EoS,
feeddown and a narrowing of the fireball rapidity distributions due to
the large transverse flow), we have reduced the total entropy in our
fireball by ca.~20\%.  With a freeze-out temperature of $T_\mathrm{fo}
\simeq 100(160) \,\MeV$, the measured $p_T$ spectra and elliptic flow of
light (multistrange) hadrons can be reasonably well described,
cf.~dashed lines in Fig.~\ref{fig_hadrons}\footnote{We have not removed
  the entropy lost to multistrange hadrons (which amounts to $\sim$2\%)
  from the fireball; we neglect this correction in both fireball and
  hydro evolution.}.  One might be concerned that the calculated
$v_2(p_T)$ for thermal pions and protons exceeds the data toward higher
$p_T$, in a regime which is still relevant for thermal photon
production. However, the thermal $p_T$ spectra start to underpredict the
experimental yields data in this regime.  For example, for pions with
$p_T\simeq3\,\GeV$, the thermal spectrum accounts for ca.~65\% of the
experimental yield;
weighting the thermal pion-$v_2$ of $\sim 22\%$ with this fraction gives
$\sim 14\%$, which is not far from the data, $v_2(p_T=3\GeV)\simeq11\%$.
While these estimates pertain to kinetic freezeout, the calculations for
the $\phi$ meson represent a snapshot at $T_{\mathrm{ch}}=160\,\MeV$;
they approximately follow the measured spectra and $v_2$ out to higher
$p_T$.

\subsection{Ideal Hydrodynamics}
\label{ssec_hydro}

The hydrodynamic model used in the present study has been described in
detail in Ref.~\cite{He:2011zx}. It is based on the 2+1-dimensional
ideal hydro code of Ref.~\cite{Kolb:2003dz} (AZHYDRO), augmented with the
updated EoS described above (\textit{latPHG}) and initial conditions tuned 
to reproduce bulk and multistrange hadron yields, spectra and $v_2$ in 
central and semicentral Au-Au collisions at full RHIC energy. Specifically, 
a rather compact initial-entropy density profile was adopted, proportional 
to the binary-collision density in the optical Glauber model (this is not
unlike what has been obtained in gluon saturation models); with a
thermalization time of $\tau_0=0.6 \; \fm/c$, the central initial
temperature amounts to $T_0=398 \;\MeV$ in 0-20\%
Au-Au($\sqrt{s_{NN}}=200\;\GeV$). Furthermore, a sizable initial radial
flow with a surface value of around $0.13 c$ has been introduced (and a
very small positive $v_2$ to optimize the hadron fits). All of these
features (lattice EoS, compact initial profile and initial radial flow)
lead to a more violent radial expansion, which enables an improved
description of the bulk-hadron $v_2$ at kinetic freezeout even within
ideal hydrodynamics.  At the same time, it generates an earlier
saturation of the bulk-medium $v_2$, which, in particular, requires
multistrange hadrons to freeze out at the chemical freezeout temperature
$T_{\rm ch}$ to reproduce their $p_T$ spectra and $v_2$ (this is not
unwelcome as their hadronic cross sections might be too small to
maintain kinetic equilibrium at lower temperatures). A more violent
expansion has also been identified as a key to the solution of the ``HBT
puzzle''~\cite{Pratt:2008qv}. As emphasized in Ref.~\cite{He:2011zx},
the ideal-hydro tunes are not meant to supplant principally more 
realistic viscous evolutions, but rather to explore limitations and
flexibilities in the (ideal-) hydro description, within ``reasonable''
variations of the input. In Fig.~\ref{fig_hadrons}, the solid lines show 
some of the hydro results for bulk spectra and elliptic flow in comparison 
to RHIC data, which turn out to be very similar to the schematic fireball.

\subsection{Comparison of Space-Time Properties}
\label{ssec_comp}
We are now in position to systematically compare the space-time
evolutions of the schematic fireball and the ideal hydro solution. We
focus on 0-20\% central Au-Au collisions at RHIC energy
($\sqrt{s}=200\,A\GeV$), where both models describe the bulk spectra and
$v_2$ fairly well, based on the same EoS (\textit{latPHG}). Since the
isotropic nature of the fireball is rather schematic compared to the
more elaborate profiles in the hydro evolution, we investigate in this
Section how this difference manifests itself in suitably averaged bulk
quantities, which are expected to play an important role in the photon
emission observables discussed in the next Section.

\begin{figure}[!t]
\begin{center}
\begin{minipage}{0.48\linewidth}
\includegraphics[width=\textwidth]{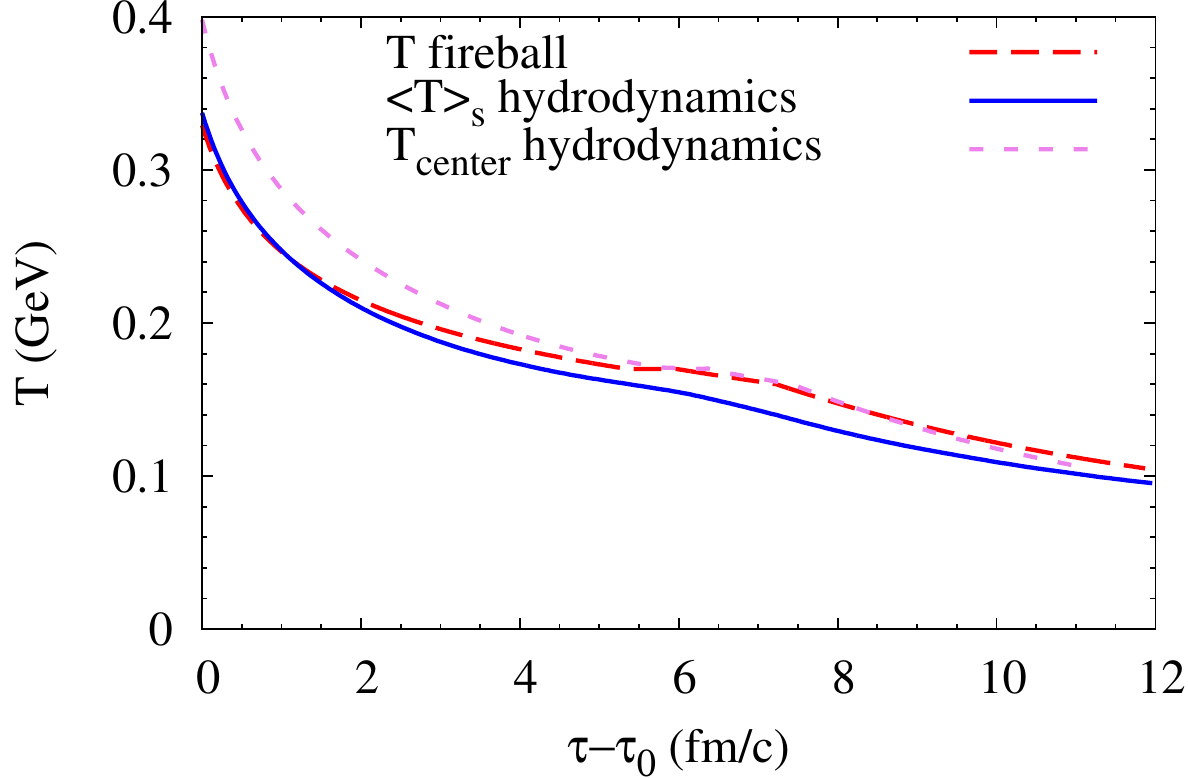}
\end{minipage}\hfill
\begin{minipage}{0.48\linewidth}
\includegraphics[width=\textwidth]{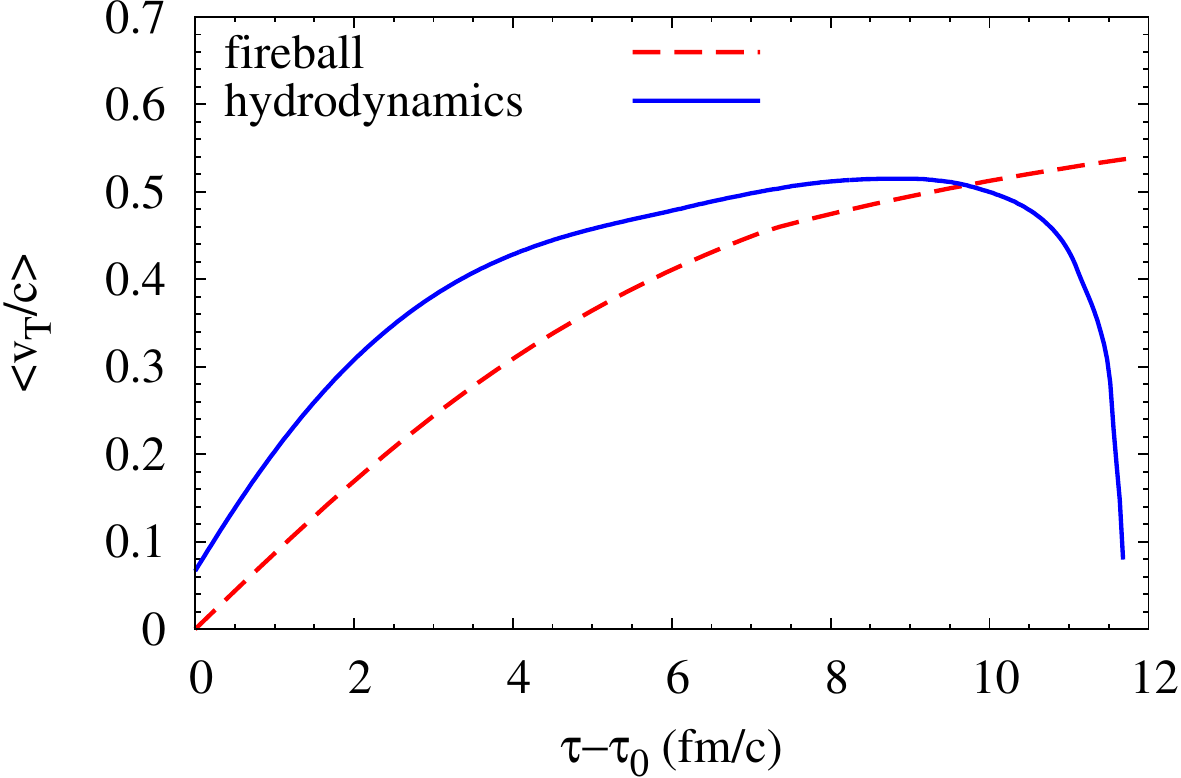}
\end{minipage}

\vspace*{5mm}
\includegraphics[width=0.48\linewidth]{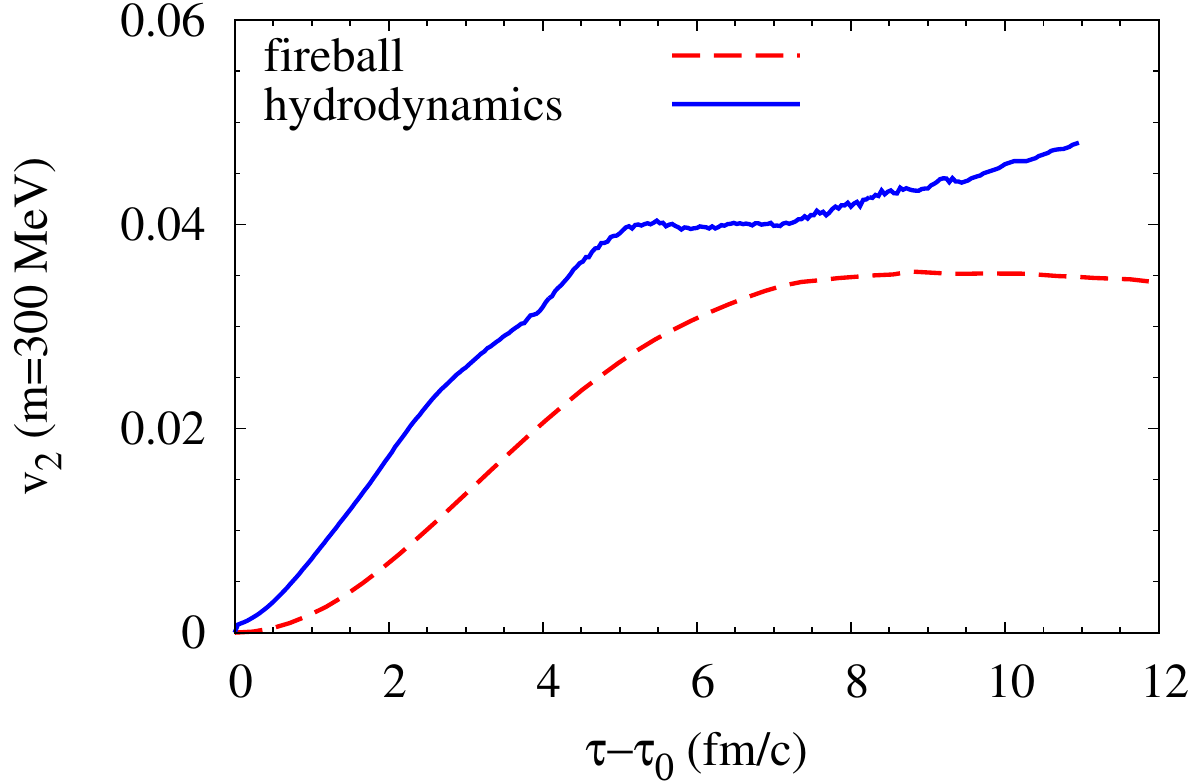}
\end{center}
\caption{(Color online) Time evolution of average temperature (upper
  left panel), average transverse flow velocity (upper right panel) and
  elliptic flow (lower panel, using a particle mass of $m=300\,\MeV$) in
  the expanding fireball (dashed lines) and hydrodynamic evolution
  (solid lines).  In addition, the upper left panel contains the
  temperature evolution of the central hydro cell (short-dashed
  line).}
\label{fig_T-eps}
\end{figure}
Let us first investigate the time evolution of (average) temperature,
radial and elliptic flow, see Fig.~\ref{fig_T-eps}. The temperature of
the fireball is generally rather close to the average one from the hydro
model, but exhibits systematically slightly higher values in the late
states of the evolution (see upper left panel). This goes along with a
10-15\% longer lifetime of the fireball evolution, indicating a somewhat
slower cooling in the later stages.  The radial flow (upper right panel 
of Fig.~\ref{fig_T-eps}) starts out higher in the hydro (due to finite
initial flow), but then levels off more quickly than in the fireball,
eventually dropping below the latter's. For the elliptic flow
comparison, we evaluate the $v_2$ coefficient of the momentum spectra of
particles with an average mass of $300\; \MeV$ at fixed proper time
(lower panel of Fig.~\ref{fig_T-eps}). For the hydrodynamic evolution
this involves a varying temperature while the fireball is spatially
homogeneous at each time.

\begin{figure}[!t]
\begin{center}
\begin{minipage}{0.48\linewidth}
\includegraphics[width=\textwidth]{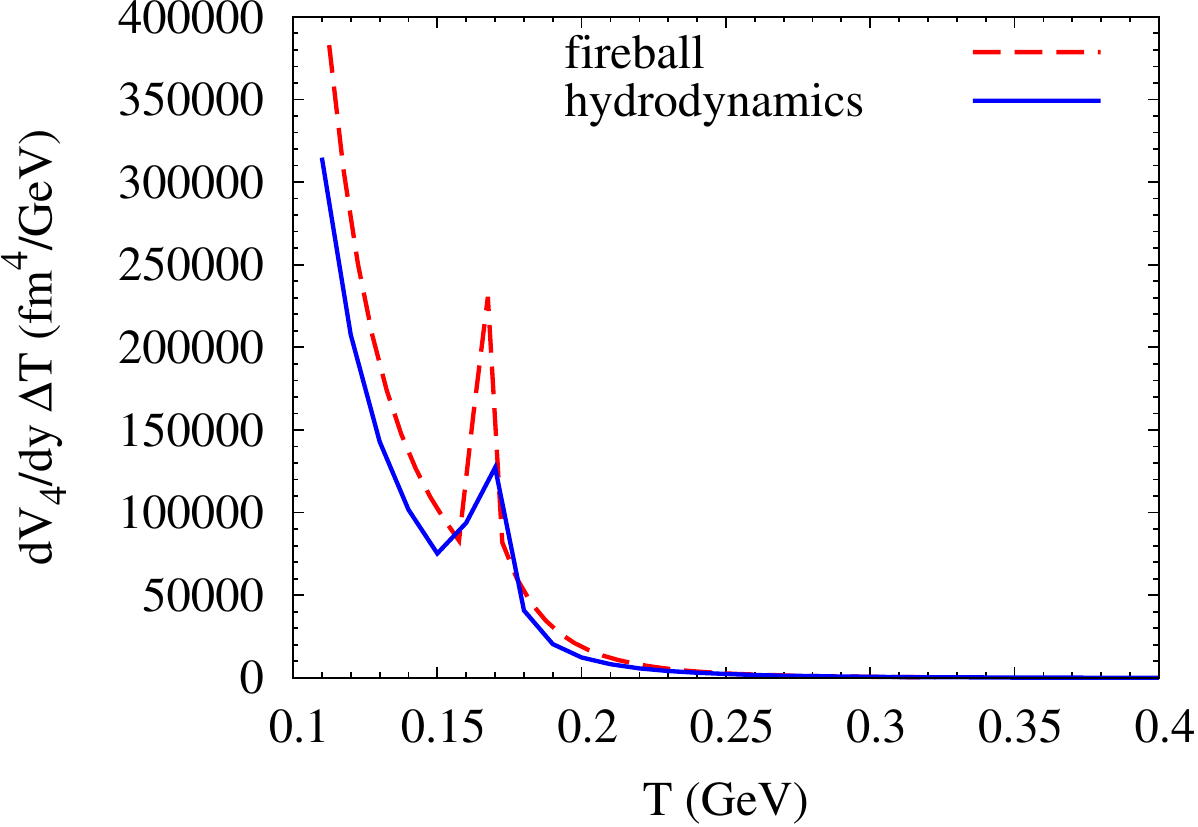}
\end{minipage}\hfill
\begin{minipage}{0.48\linewidth}

\vspace*{5mm}
\includegraphics[width=\textwidth]{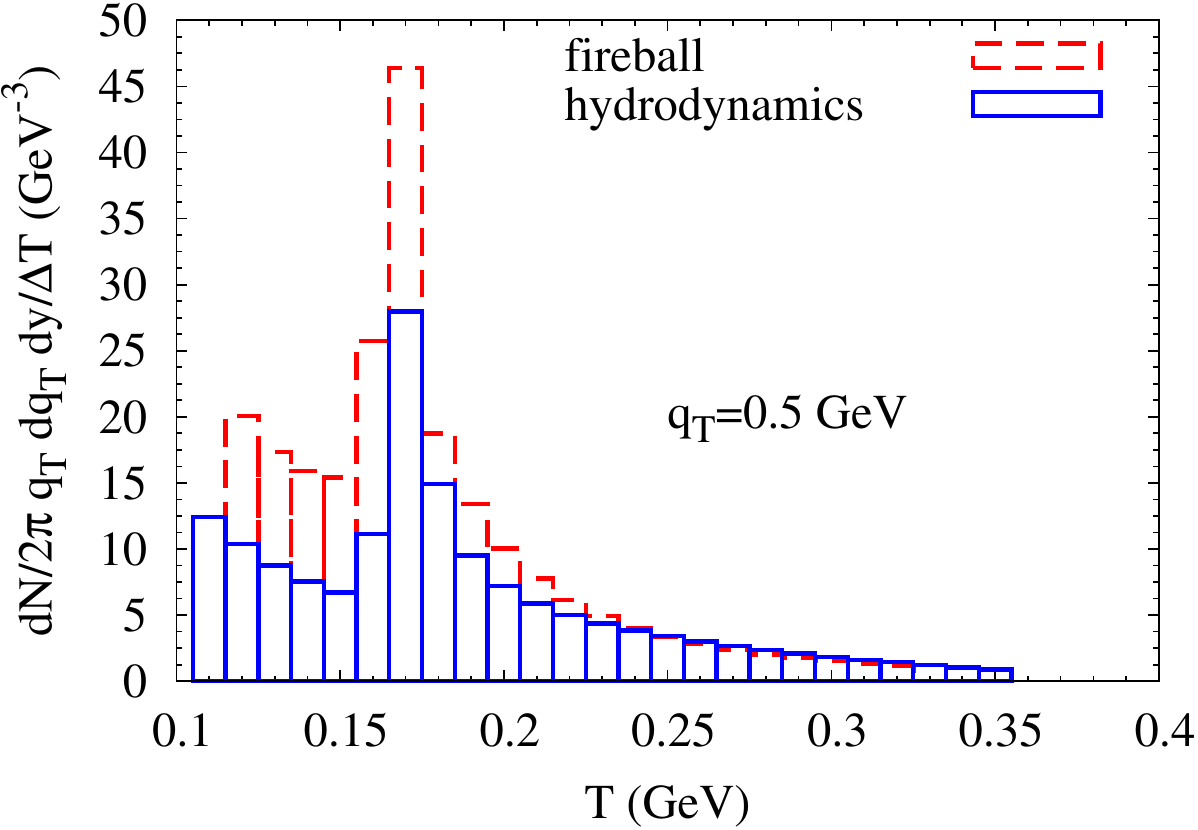}
\end{minipage}
\includegraphics[width=0.48 \linewidth]{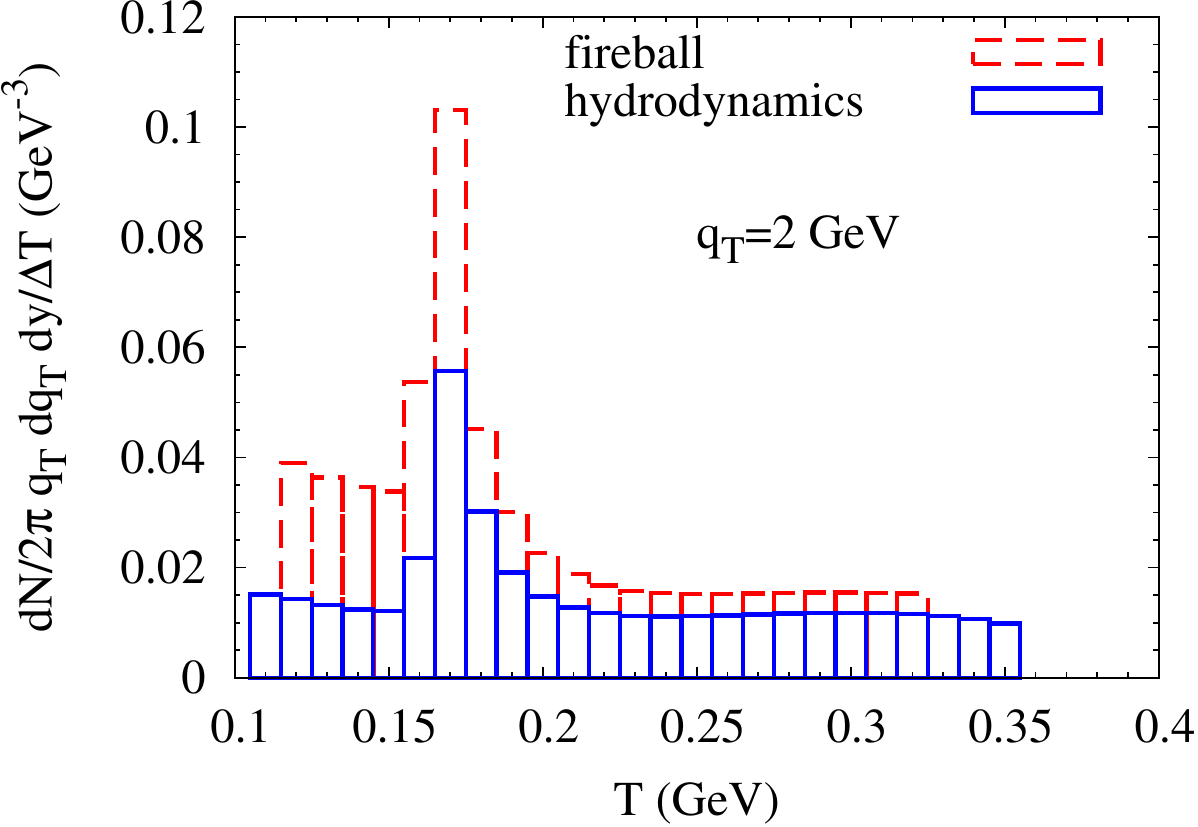}
\end{center}
\caption{(Color online) Temperature evolution of the differential
  emission four-volume (upper left panel) and the double-differential
  photon emission rate (QGP for $T>T_\mathrm{pc}$ and hadronic for
  $T<T_\mathrm{pc}$) for two transverse momenta ($q_T=0.5 \, \GeV$ and
  $q_T=2 \, \GeV$ in the upper right and lower panel, respectively), in
  the expanding fireball (dashed lines) and hydrodynamic evolution
  (solid lines).}
\label{fig_V4-spec}
\end{figure}
\begin{figure}[!t]
\centering{\includegraphics[width=0.48\linewidth]{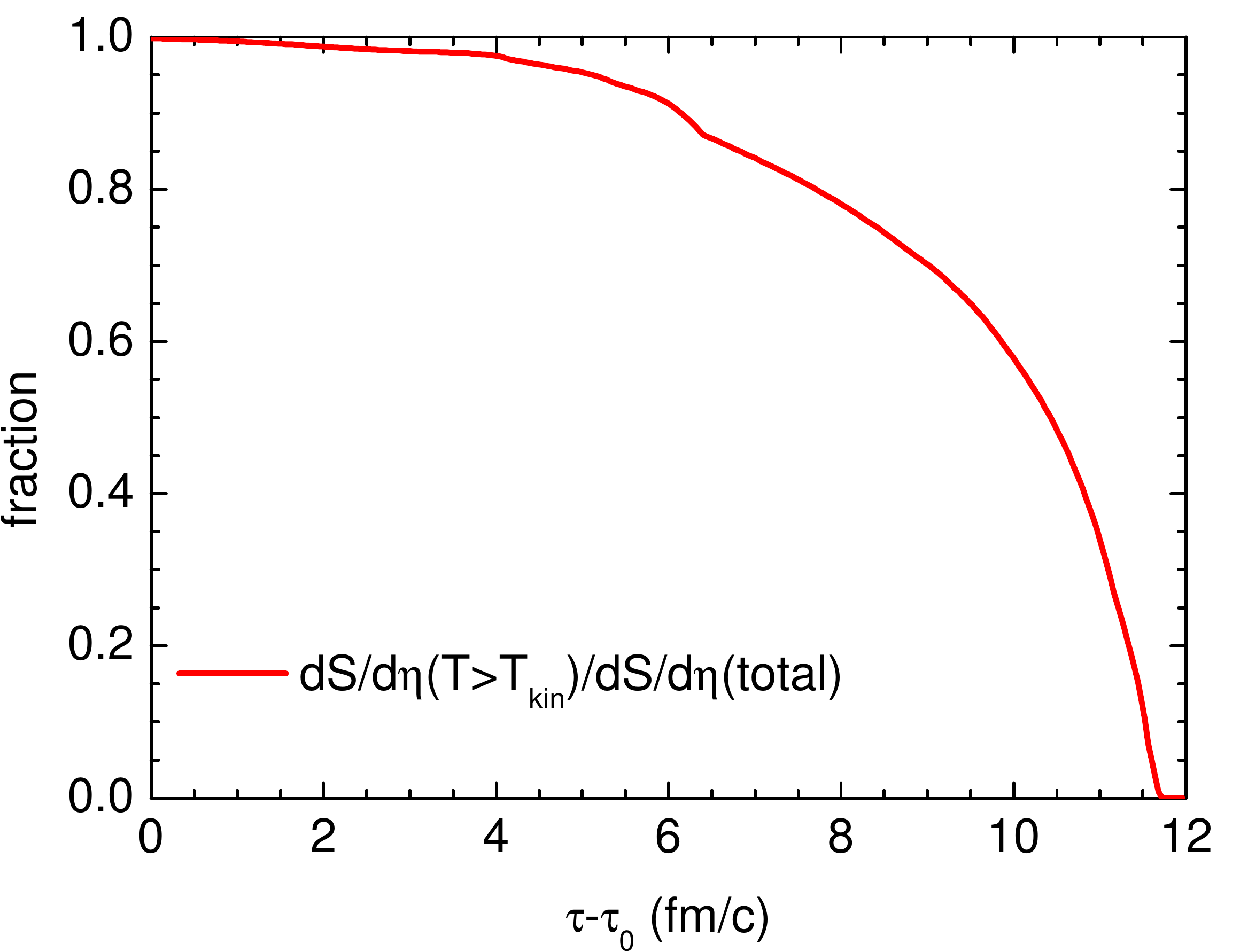}}
\caption{(Color online) Time evolution of the entropy fraction (relative
  to the total) which is in fluid cells at temperatures above the
  kinetic freezeout temperature in the hydrodynamic model.}
\label{fig_entro}
\end{figure}

To properly interpret the observed photon spectra it is important to
understand how the different emission stages contribute to the total.
From the rate expression, Eq.~(\ref{rate}), one sees that the weighting
is governed by three ingredients: the (differential) four-volume, the
thermal weight (Bose factor) and the EM spectral function (at the photon
point). The former two are governed by the bulk-medium evolution. To
make a closer connection to the underlying matter properties, we now
plot pertinent quantities as a function of temperature, rather than
time. The upper left panel in Fig.~\ref{fig_V4-spec} shows the $T$ 
dependence of the differential four-volume, $\Delta V_4/\Delta T$, 
over temperature intervals of $\Delta T=10\,\MeV$ (and per unit 
rapidity). For both fireball and hydro evolution this quantity shows 
a distinct maximum structure around the pseudocritical temperature of
$T_\mathrm{pc}\simeq170 \,\MeV$, as a consequence of the rapid change in
the entropy density in the EoS (a remnant of the latent heat). One also
finds a pronounced increase in the differential four-volume in the
late(r) hadronic stages of the collision (stipulating the importance of
a realistic hadronic photon emission rate). Again we find that the ideal
hydro evolution seems to cool somewhat faster than the fireball (this
might slightly change in a viscous evolution where some of the expansion
work is dissipated into heat). Whereas hadron emission at kinetic
freezeout in the hydro evolution is a continuous process, the fireball
freezeout of the entire three-volume occurs at the end of the
evolution. This difference is illustrated in Fig.~\ref{fig_entro}, where
we plot the {\em time} dependence of the fraction of the total entropy
that is above the kinetic-freezeout temperature in the hydro
evolution. This fraction shows a marked departure from one at times
already well before the total lifetime; in contrast, this fraction is
equal to one throughout the fireball evolution.  Recall, however, that
the three-volume at small temperatures must be very similar in both
evolutions, since the total three-volume at freeze-out figures into the
calculation of hadron multiplicities (and spectra), which agree rather
well. Our hydro evolution, on the other hand, does not allow for the
possibility that the freeze-out front ``re-swallows'' previously
frozen-out matter cells. This effect has been studied, \eg, in
Ref.~\cite{Grassi:2004dz}, where it was found to be significant even in
the context of hadron observables. It would be interesting to
investigate its impact on thermal-photon spectra.

The increase in emission four-volume is counteracted by the drop in
temperature, suppressing the thermal distribution function in the rate. 
This renders the energy argument in the Bose function as an important 
scale. As is well known (see, \eg, Ref.~\cite{Rapp:2011is} for an 
analogous study in the dilepton context), larger energies increase the 
sensitivity of the exponential to temperature variations and thus will 
lead to a stronger weighting of earlier phases. To exhibit this interplay 
in a more realistic way, we include the weights from the QGP and hadronic 
spectral functions, $\im \Pi_\mathrm{EM}$, in plotting the 
temperature-differential photon spectra for two representative transverse 
energies (see upper right and lower panel of Fig.~\ref{fig_V4-spec}); in
other words, we use the full rate expression - the same for both 
evolution models - figuring in our comparisons to data in the following
sections (recall that the AMY QGP rates (full LO
result)~\cite{Arnold:2001ms} and the TRG hadronic
rates~\cite{Turbide:2003si} are nearly degenerate around
$T_\mathrm{pc}$, thus avoiding a ``bias'' for either phase). For low
photon momenta (energies), $q_T=0.5\,\GeV$, the ``phase transition''
peak observed in the four-volume is remarkably enhanced, but also the
high-temperature part now exhibits significant emission strength. As
expected, the high-temperature component increases further at larger
momentum ($q_T =2\, \GeV$), albeit not dramatically. A pronounced peak
around $T_\mathrm{pc}$ persists also for these kinematics.
  
This analysis clearly identifies the importance of the
``pseudo-critical'' regime, around $T_\mathrm{pc}$, for thermal photon
radiation, as found in Ref.~\cite{vanHees:2011vb}. The macrophysics
encoded in the underlying EoS plays an important role through a rapid
change in entropy density over a rather small temperature window,
possibly augmented by a reduction in the velocity of sound,
$c_{\mathrm{s}}^2$, figuring into the hydro evolution (a pertinent
slowing down has not been implemented into the fireball evolution). An
equally important role is plaid by the microphysics, \ie, relatively
large hadronic emission rates, comparable to the QGP ones, in the
transition region. Together with substantial flow-induced blue shifts,
this led to generating a sizable photon $v_2$ in our previous fireball
calculations~\cite{vanHees:2011vb}.  These features can be recovered
within hydrodynamic evolutions, \emph{if} the collective flow is built
up sufficiently fast, which can be realized via a modest initial
radial-flow field and a compact initial-density profile. However,
quantitative differences remain, which we further analyze in the
following two sections by comparing the photon spectra from both
approaches with each other and to direct photon data at RHIC and LHC.

\section{Direct Photon Spectra at RHIC and LHC}
\label{sec_gam}
All thermal photon spectra presented in this Section are based on the
same emission rates, from a complete LO calculation in a perturbative
QGP~\cite{Arnold:2001ms} and hadronic many-body calculations
supplemented with $t$-channel meson exchange 
reactions~\cite{Turbide:2003si} as well as $\pi\pi$ and $\pi K$
Bremsstrahlung~\cite{Liu:2007zzw}. We also note that short-lived (strong) 
resonance decays are usually not subtracted from the experimental spectra. 
To account for these ``strong feeddown'' photons (\eg, $\Delta\to N\gamma$, 
etc.), we follow Refs.~\cite{Rapp:1999us,vanHees:2011vb} by running our
evolution for an extra $1\,\fm/c$ after kinetic freezeout. Strictly speaking, 
as elaborated in Ref.~\cite{vanHees:2007th}, these final-state decays would 
have to be calculated with slightly modified kinematics compared to thermal
processes (with an extra Lorentz-$\gamma$ factor), but we neglect this
difference in the present study. Ultimately, their contribution to the total 
direct-$\gamma$ $v_2$ turns out to be rather modest (e.g., increasing it 
by typically 5-10\% around $q_T=2$\,GeV), and even less for the spectra.

\subsection{Thermal Fireball}
\label{ssec_gam-fb}
\begin{figure}[!t]
\centering{\includegraphics[width=0.48 \linewidth]{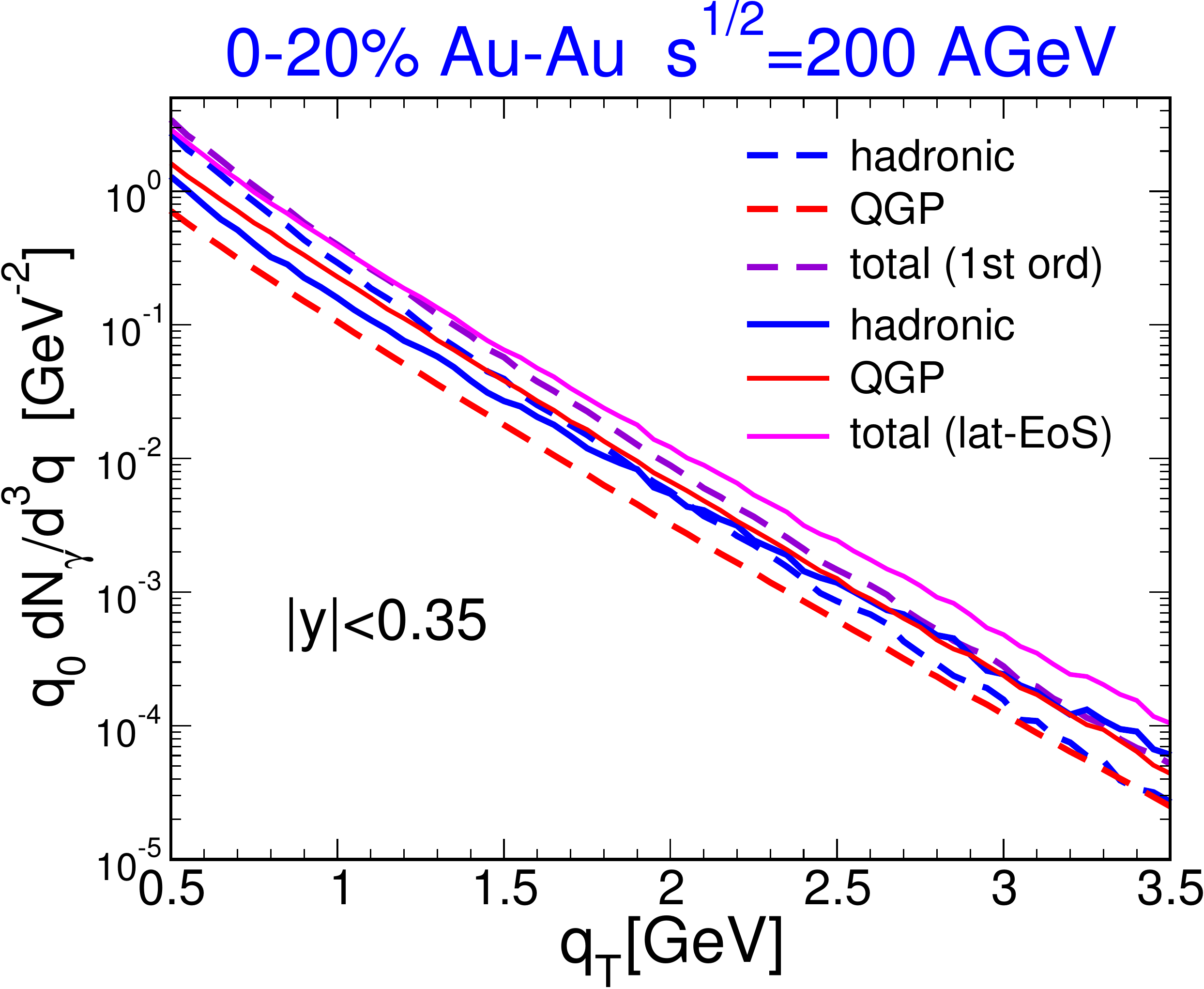}}
\caption{(Color online) Comparison of thermal photon spectra in
  Au+Au($\sqrt{s}=0.2A \, \TeV$) collisions from an expanding fireball
  using either a first-order quasiparticle-QGP + HRG EoS with
  $T_\mathrm{c}=T_\mathrm{cm}=180\,\MeV$~\cite{vanHees:2011vb} (dashed
  lines) or a cross-over lQCD + HRG EoS with $T_\mathrm{pc}=170\, \MeV$
  and $T_\mathrm{ch}=160\, \MeV$; red, blue and purple/pink lines
  represent QGP, hadronic and total contributions, respectively.}
\label{fig_fb-rhic-comp}
\end{figure}

We start by updating the fireball calculations at RHIC, which in
Ref.~\cite{vanHees:2011vb} where conducted with a first-order EoS (and a
by now outdated chemical-freezeout temperature of $T_\mathrm{ch}=180\,
\MeV$), by implementing the \textit{latPHG} EoS of Ref.~\cite{He:2011zx}
with chemical freezeout at $T_\mathrm{ch}=160\, \MeV$.  The resulting
thermal spectra are compared to the results of
Ref.~\cite{vanHees:2011vb} in Fig.~\ref{fig_fb-rhic-comp} using the same
total entropy. Since the partitioning of hadronic and QGP emission in
the mixed phase of the first-order transition is now entirely assigned
to the nonperturbative QGP phase, and due to a lower critical
temperature in the \textit{latPHG}, the QGP contribution significantly
increases while the hadronic part decreases compared to the first-order
scenario. This ``reshuffling'' by itself corroborates that the major
portion of the thermal emission originates from around the phase
transition region, independent of the details of the EoS. While the
total (integrated) photon yield is not changed much (analogous to what
has been found for low-mass dileptons~\cite{Rapp:2013nxa}), the
high-$q_T$ part of the spectrum benefits from the increased temperature
in the QGP close to $T_\mathrm{pc}$ and from the later chemical
freezeout in the hadronic phase, which slows down the drop in
temperature that arises in the presence of pion (and other effective)
chemical potentials (for the inclusive yields, and at low $q_T$, the
faster temperature drop is (over-) compensated by the fugacity factors).

\begin{figure}[!t]
\begin{minipage}{0.48\linewidth}
\includegraphics[width=\textwidth]{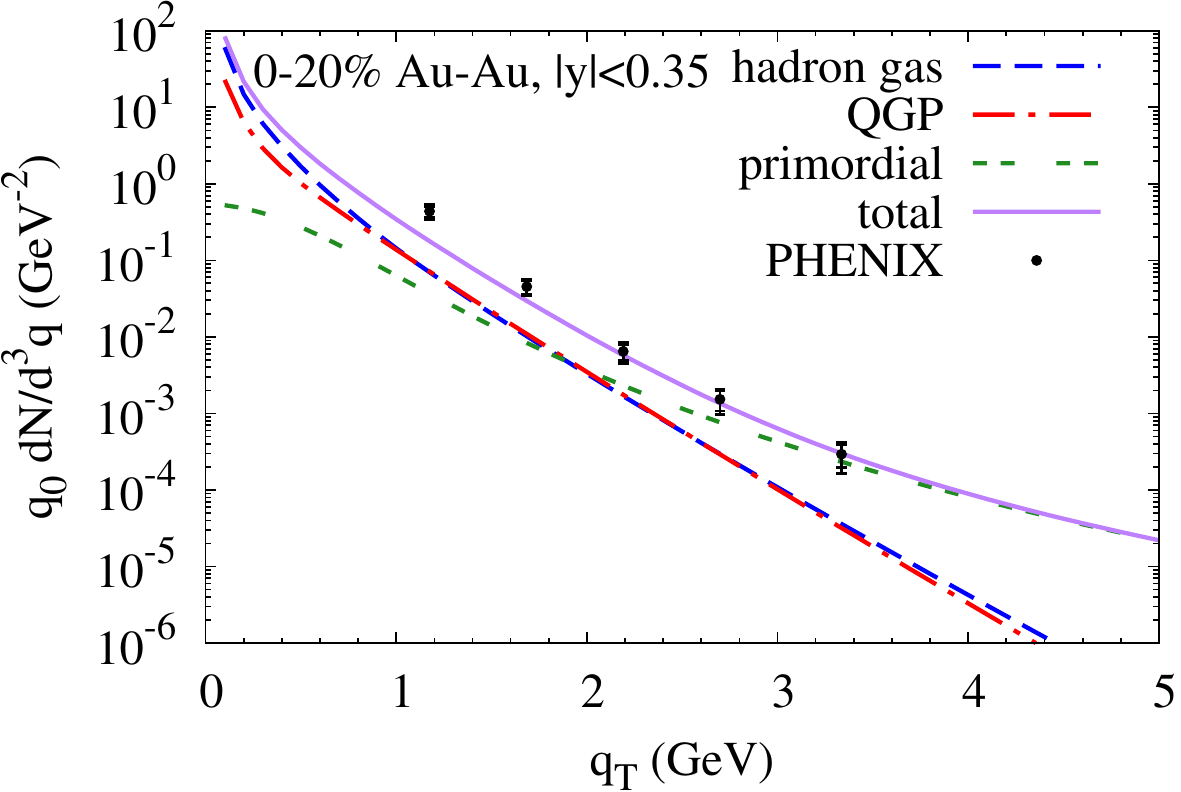}
\end{minipage}\hfill
\begin{minipage}{0.48\linewidth}
\includegraphics[width=\textwidth]{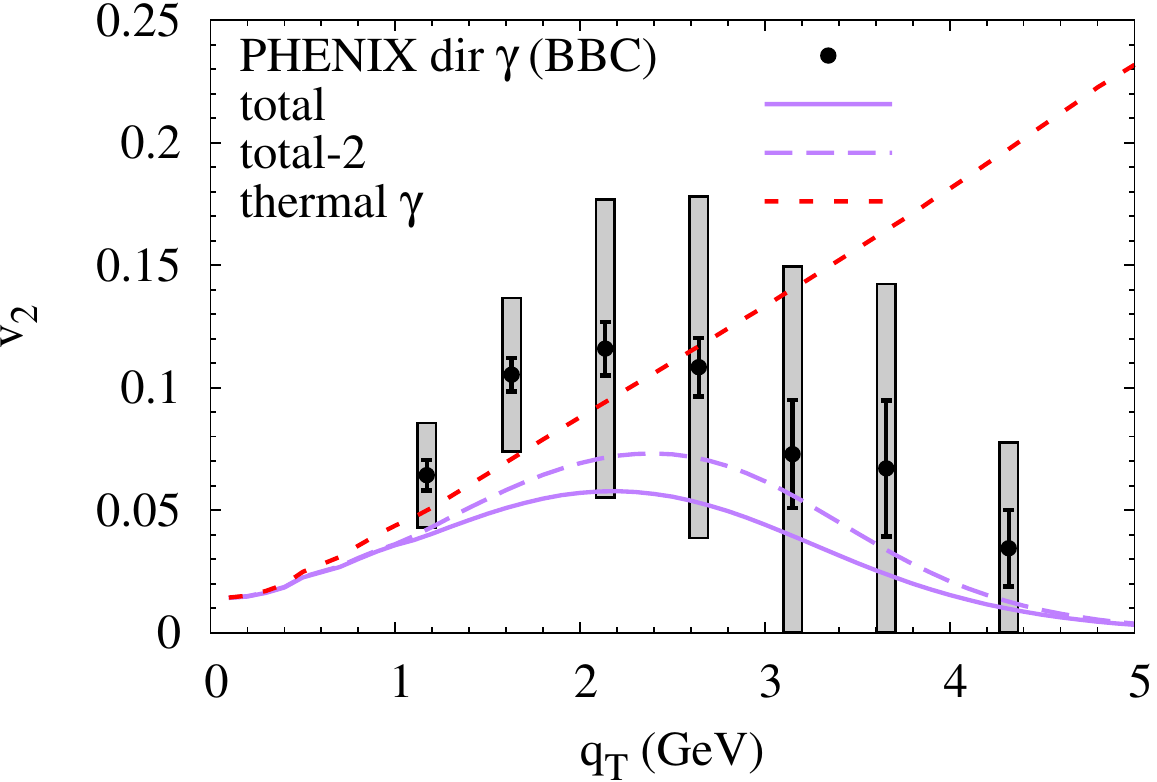}
\end{minipage}
\caption{(Color online) Direct photon spectra (left panel) and elliptic
  flow (right panel) from the expanding fireball in 0-20\%
  Au+Au($\sqrt{s}=0.2\,A \TeV$) collisions with updated total entropy
  using the \textit{latPHG} EoS, compared to PHENIX
  data~\cite{Adare:2008fq,Adare:2011zr}. In the left panel, blue dashed,
  red dashed-dotted, green short-dashed-dotted and purple solid lines
  correspond to hadronic, QGP and primordial contributions, and their
  sum (``total''), respectively. The primordial conribution is based on
  the PHENIX pp parameterization. In the right panel, the red
  short-dashed line is the combined thermal $v_2$; the purple solid and
  long-dashed lines are the total direct-photon $v_2$ using two
  different primordial contributions, either the PHENIX pp
  parameterization (as in the left panel) or an $x_t$-scaling ansatz
  (labeled ``total-2'').  Both primoridial contributions are assumed to
  carry vanishing $v_2$.}
\label{fig_fb-rhic}
\end{figure}
\begin{figure}[!t]
\begin{minipage}{0.48\linewidth}
\includegraphics[width=\textwidth]{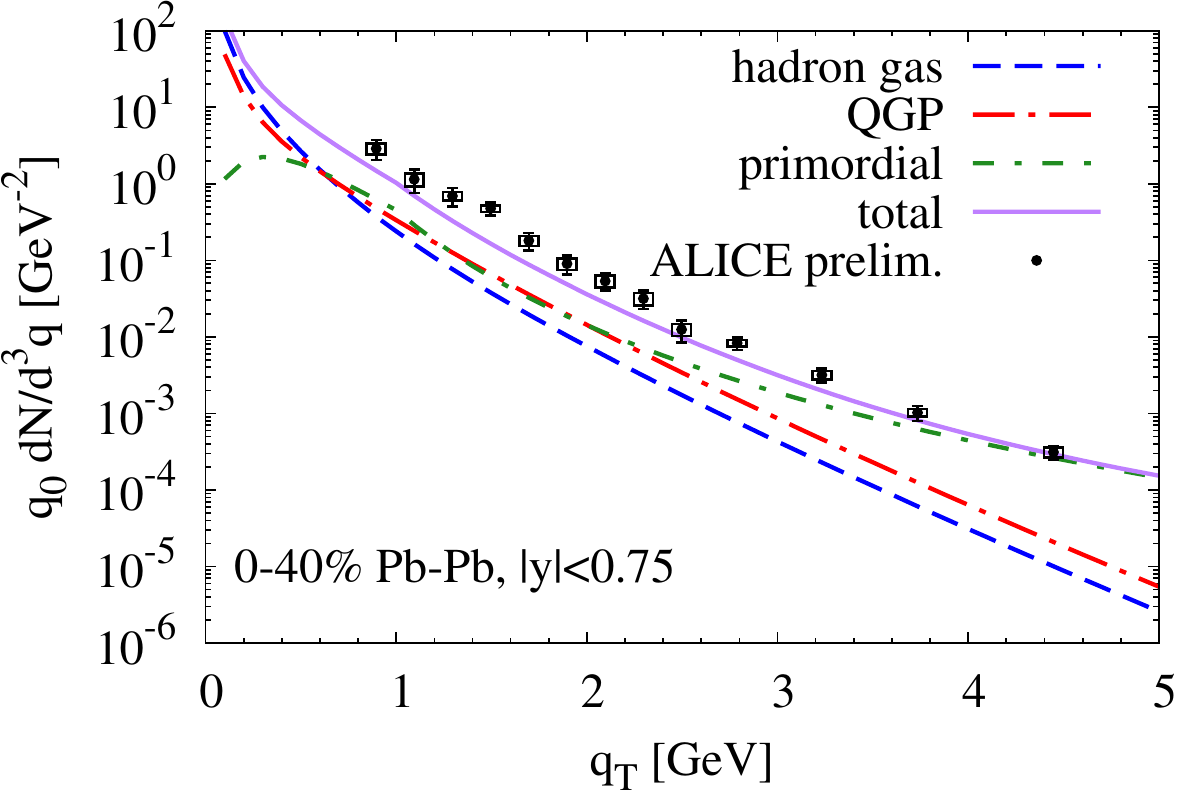}
\end{minipage}\hfill
\begin{minipage}{0.48\linewidth}
\includegraphics[width=\textwidth]{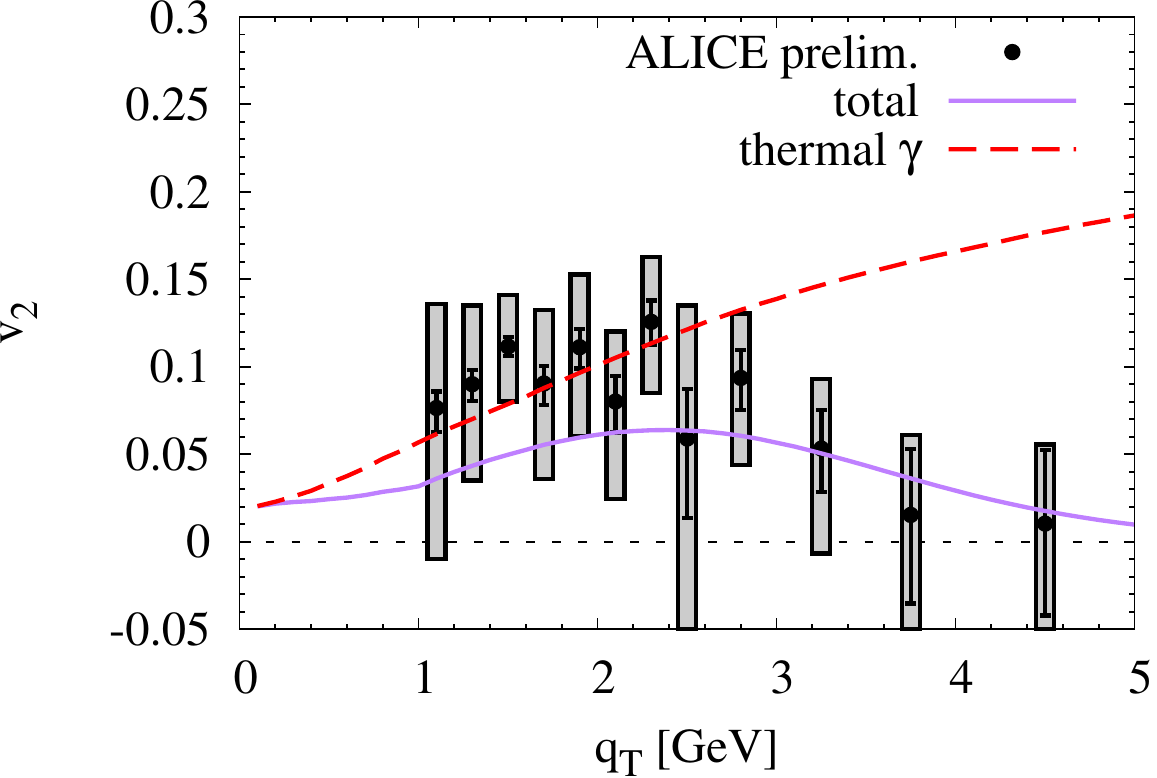}
\end{minipage}
\caption{(Color online) Direct photon spectra (left panel) and elliptic
  flow (right panel) from the expanding fireball in 0-40\%
  Pb+Pb($\sqrt{s}=2.76\,A\TeV$) collisions using \textit{latPHG} EoS,
  compared to preliminary ALICE data~\cite{Wilde:2012wc,Lohner:2012ct}.
  In the left panel, red, blue, green and purple lines represent QGP,
  hadronic and primordial contributions, as well as the total, respectively. 
  In the right panel we show the combined thermal $v_2$ (red dashed line) and
  the total $v_2$ (purple solid line).}
\label{fig_fb-lhc}
\end{figure}

As mentioned above, we have further updated our fireball calculations by
a careful readjustment of the entropy when using \textit{latPHG}, leading 
to a 20\% decrease compared to Ref.~\cite{vanHees:2011vb}. With our nonlinear 
dependence of thermal photon production on the charged-particle multiplicity, 
$\propto N_\mathrm{ch}^x$ with $x \simeq 1.5$~\cite{Turbide:2003si,Rapp:2013nxa} 
(larger at higher $q_T$), the thermal yields are somewhat reduced compared to 
our earlier results. The comparison to the PHENIX data for direct photons in
Fig.~\ref{fig_fb-rhic} shows discrepancies at low $q_T$ for both
spectral yields and $v_2$, while for $q_T \ge 2 \, \GeV$ the
calculations are within the experimental errors. We illustrate
uncertainties in the determination of the primordial photon component
due to initial hard $NN$ scatterings: on the one hand, we used a
phenomenological parameterization by the PHENIX collaboration of their pp
data~\cite{Adare:2008fq}; on the other hand, we used the $x_t$-scaling
ansatz of Ref.~\cite{Srivastava:2001bw}, fitted to the high-$q_T$ part
of the PHENIX pp data with a $K$-factor of $2.5$.  The latter spectrum
turns out to be somewhat smaller than the PHENIX parameterization at
small $q_T$. This has rather little impact on the total $q_T$ spectrum,
but it affects the $v_2$ more significantly, inducing an increase of the
total direct-photon $v_2$ of up to $\sim 25\%$ around $q_T\simeq2.5 \,
\GeV$. Further theoretical studies of the hard component are needed to
better quantify this effect, \eg, via a suppression of fragmentation
photons or nuclear effects on the initial parton distribution functions.

Finally, we turn to LHC energies, where preliminary direct-photon data
are available from ALICE for 0-40\% central Pb+Pb($\sqrt{s}=2.76\,A
\TeV$) collisions~\cite{Wilde:2012wc,Lohner:2012ct}. We model these
reactions with an average charged-particle multiplicity of $\dd
N_\mathrm{ch}/\dd y=1040$ over a rapidity interval of $|y|<0.75$. For
primordial photon production from binary $NN$ collisions, we employ the
$x_t$-scaling ansatz~\cite{Srivastava:2001bw}, fitted to the high-$q_T$
ALICE photon spectra with a $K$-factor of 2. The description of the
spectra and $v_2$ is at a comparable level as at RHIC, with indications
for an underestimate in both observables in the regime where thermal
radiation is most significant, cf.~Fig.~\ref{fig_fb-lhc}.

\subsection{Ideal Hydrodynamics}
\label{ssec_gam-hydro}

\begin{figure}[!t]
\begin{minipage}{0.48\linewidth}
\includegraphics[width=\textwidth]{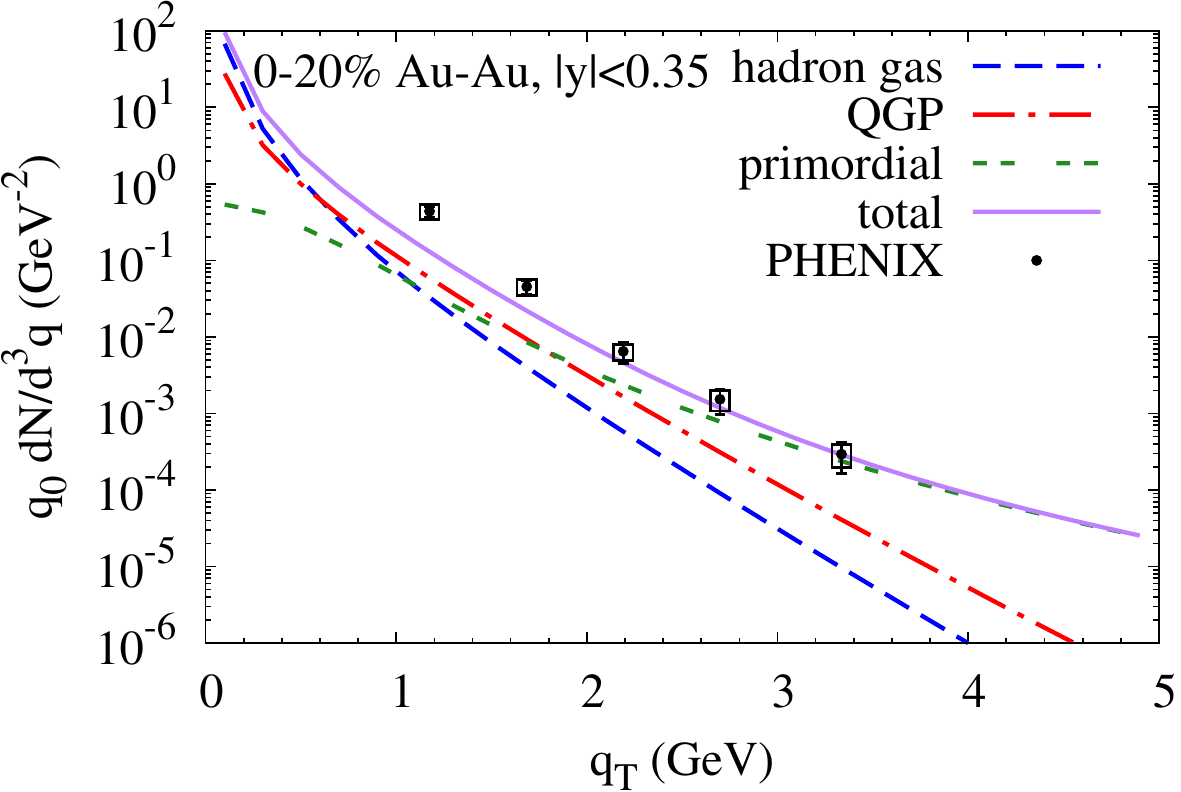}
\end{minipage}\hfill
\begin{minipage}{0.48\linewidth}
\includegraphics[width=\textwidth]{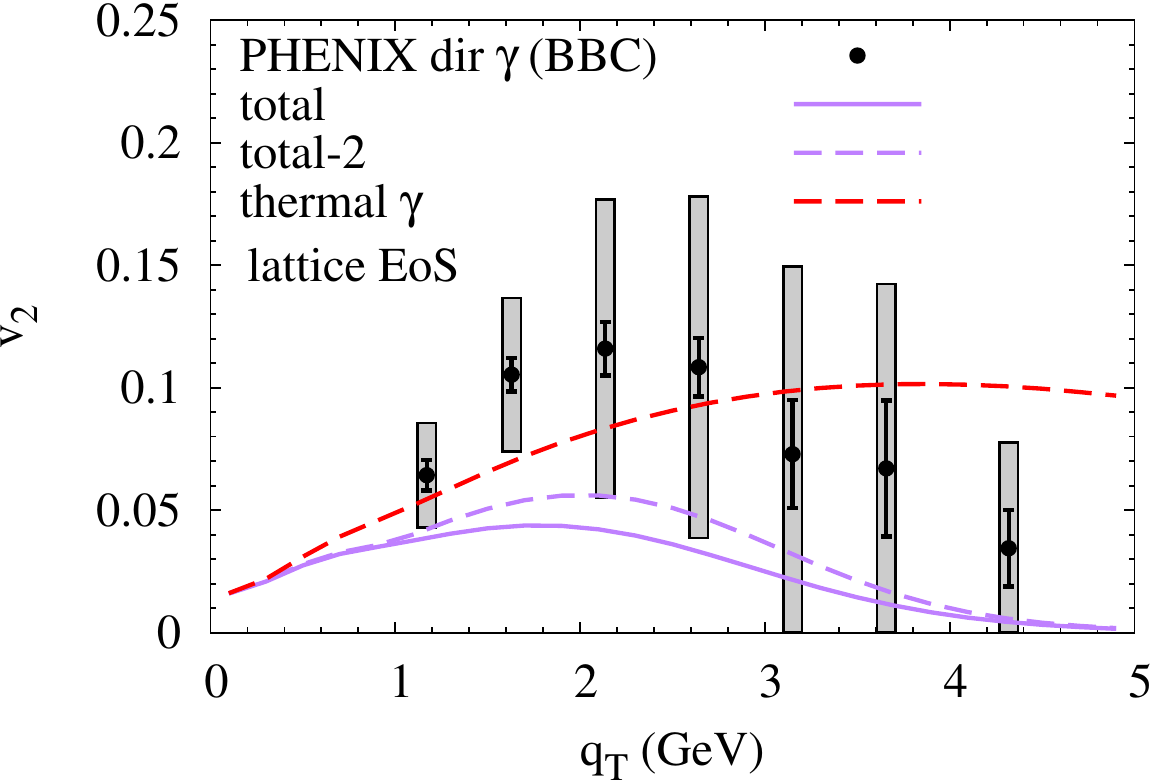}
\end{minipage}
\caption{(Color online) Direct photon spectra (left panel) and $v_2$
  (right panel) in Au+Au($\sqrt{s}=0.2\, A\TeV$) using ideal
  hydrodynamics, compared to PHENIX
  data~\cite{Adare:2008fq,Adare:2011zr}; line identifications as
  Fig.~\ref{fig_fb-rhic}.}
\label{fig_hy-rhic}
\end{figure}

The direct-photon results from the ideal-hydro tune with \textit{latPHG}
EoS and default emission rates at RHIC are displayed in
Fig.~\ref{fig_hy-rhic}. The QGP contribution to the $q_T$ spectra agrees
within ca.~30\% with the fireball results, but the hadronic portion
falls short by a larger margin, especially toward higher $q_T$. This is
not unexpected as the lifetime of the hydro evolution in the hadronic
phase is noticeably smaller, due to a faster cooling in the local rest
frame of the ideal hydro cells (leading to smaller four-volumes, recall
Fig.~\ref{fig_V4-spec}). In addition, the average temperature in the
late stages is smaller in hydrodynamics than in the fireball (cf.~upper
left panel of Fig.~\ref{fig_T-eps}), which leads to a reduction especially in
the high-$q_T$ region of the hadronic emission. Consequently, the hydro
spectra and $v_2$ come out lower than for the fireball evolution; this
increases the discrepancy with the PHENIX data for both observables,
although not by much. Both evolution models result in an underestimate
of the first two data points of the PHENIX spectra, and barely reach
into the lower portions of the error bars of the $v_2$ data: the maximal
$v_2$ reaches $\sim 4.4\%$ for the hydrodynamic evolution, compared to
$\sim 5.7\%$ for the fireball, both when using the PHENIX pp baseline
spectra (larger for the $x_t$ scaling ansatz).  However, our hydro
results are well above other hydrodynamic calculations reported in the
literature~\cite{Holopainen:2011pd,Dion:2011pp,Shen:2013vja}.  One
difference lies in a faster build-up of the $v_2$, which essentially
saturates when the system reaches the pseudo-critical region in the
cooling process, for both fireball and hydro (cf.~lower panel of
Fig.~\ref{fig_T-eps}). As mentioned above, this feature is essential in
describing the spectra and $v_2$ of multistrange hadrons with an early
kinetic freezeout close to the chemical freezeout temperature, and thus
rather well motivated by hadron phenomenology (including the
constituent-quark number scaling of $v_2$). In the hydrodynamic modeling
this can be realized by initial conditions including a finite transverse
flow at thermalization, together with a compact energy-density profile.
Both features increase the transverse flow early on, which ultimately
leads to an earlier saturation of $v_2$ (since the initial spatial
anisotropy is converted faster into momentum space); it also increases
the blue shift of the photons emitted from around $T_\mathrm{pc}$, which
helps to build up the photon yield with large $v_2$ in the $q_T=2$-$3\,
\GeV$ region. Another difference to existing hydro calculations is the
larger rate in the hadronic phase, in particular the contributions
associated with the photon point of the in-medium $\rho$ spectral
function, which includes sources from interactions involving baryons and
antibaryons~\cite{Rapp:1999us} and higher excited meson
resonances~\cite{Rapp:1999qu}.

\begin{figure}[!t]
\begin{minipage}{0.48\linewidth}
\includegraphics[width=\textwidth]{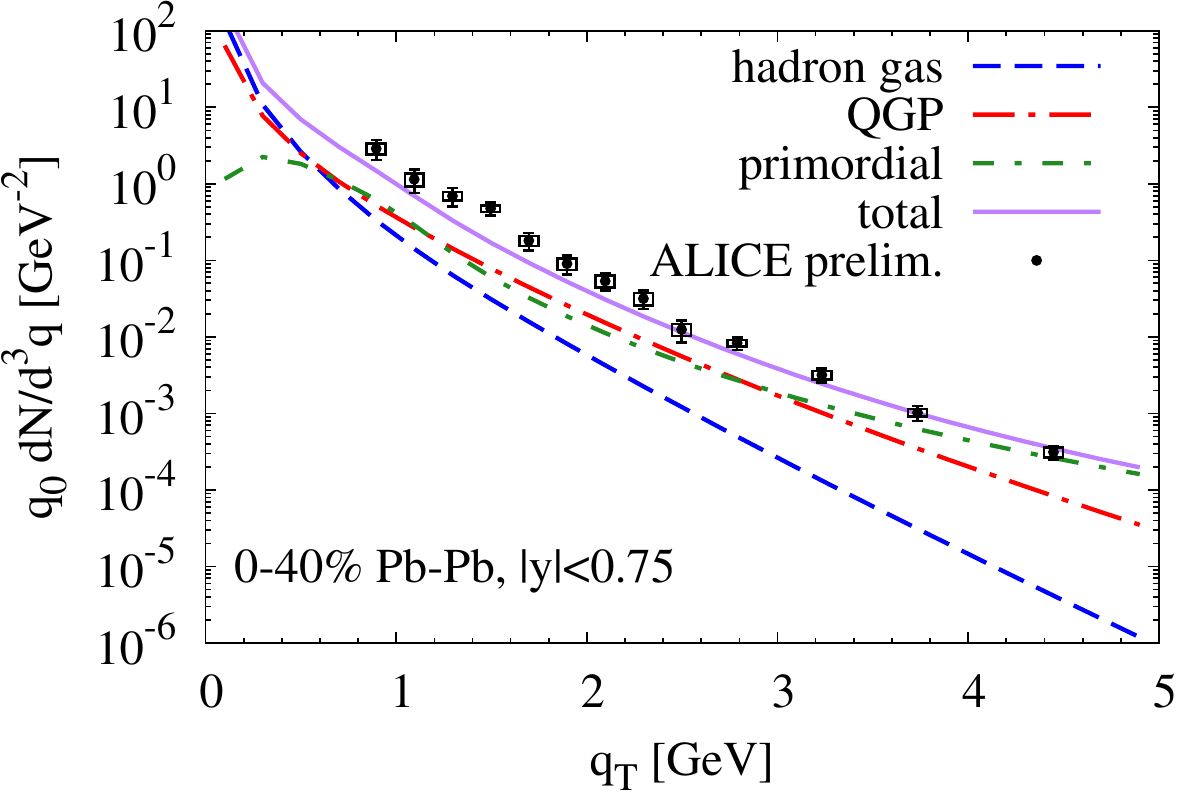}
\end{minipage}\hfill
\begin{minipage}{0.48\linewidth}
\includegraphics[width=\textwidth]{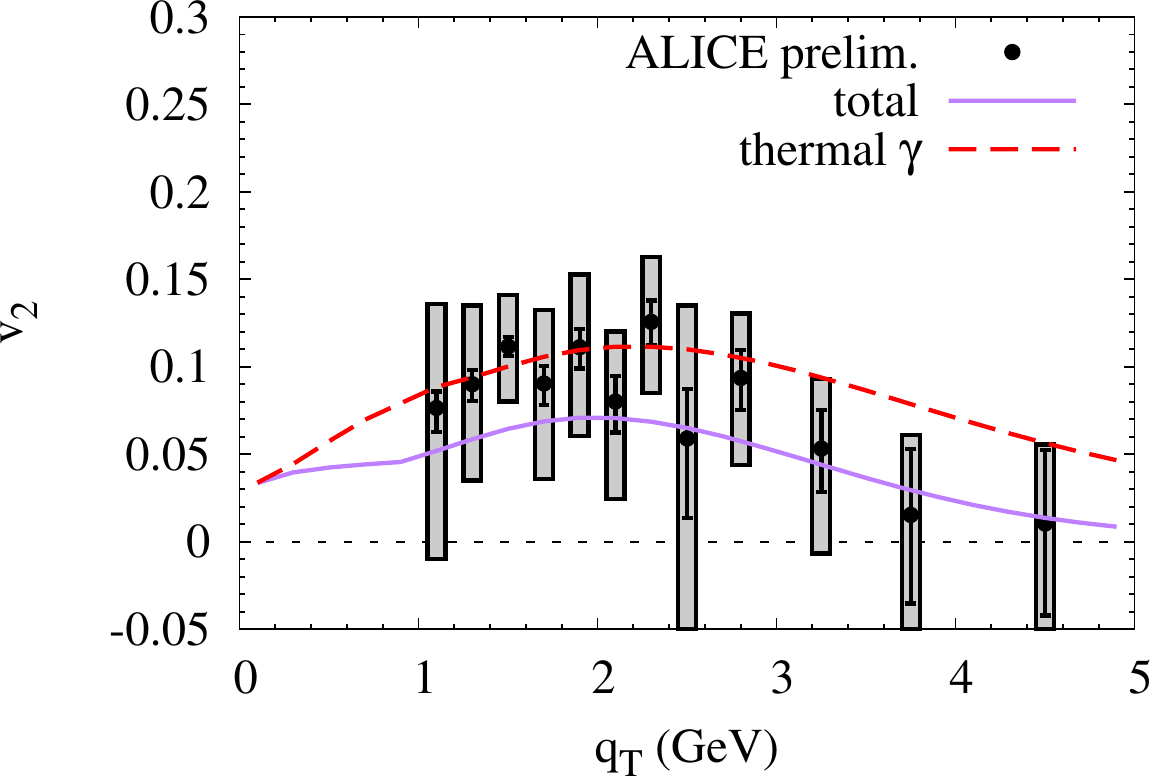}
\end{minipage}
\caption{(Color online) Direct photon spectra (left panel) and $v_2$
  (right panel) in 0-40\% Pb+Pb($\sqrt{s}=2.76\,A \TeV$) using ideal
  hydrodynamics, compared to preliminary ALICE
  data~\cite{Wilde:2012wc,Lohner:2012ct}; line identifications as in
  Fig.~\ref{fig_hy-rhic}.}
\label{fig_hy-lhc}
\end{figure}

Next, we turn to our hydro results at LHC, adopting a similar ansatz for
the initial conditions as at RHIC, \ie, with a finite transverse flow
and compact entropy density profile. A factor of $\sim 2$ underestimate of
the preliminary photon spectra measured by ALICE in 0-40\%
Pb-Pb($\sqrt{s} =2.76\, A\TeV$) is found for momenta below $q_T \simeq
2\, \GeV$, see left panel of Fig.~\ref{fig_hy-lhc}. The calculated $v_2$ is 
not inconsistent with these data within the current experimental 
uncertainties, see right panel of Fig.~\ref{fig_hy-lhc}. Although the hadronic
component in the thermal spectra is again considerably smaller than in
the fireball spectra, the relatively stronger QGP component in both
calculations compared to RHIC renders the overall impact of the hadronic
emission less relevant. The QGP component from the hydro is now somewhat
stronger than from the fireball (especially at high $q_T$), leading to
closer agreement in the total photon spectra and $v_2$, and also with
the data.  Note that the $v_2$ of the thermal component at
$q_T\gsim3\,\GeV$ is significantly larger for the fireball than for the
hydro (cf.~dashed lines in the right  panels of Figs.~\ref{fig_fb-lhc}
and \ref{fig_hy-lhc}, respectively), due to the larger hadronic
component in the former.

\section{Discussion}
\label{sec_disc}

The general trend of our results reported above for both fireball and
hydrodynamic evolutions is an underestimate of both spectra and $v_2$
for both PHENIX and preliminary ALICE data for photon momenta $q_T \le
3\, \GeV$, which is the region where the thermal radiation is
expected to be most relevant. In the following, we will investigate
possibilities how these deficits may be overcome. For simplicity, we
concentrate on the hydrodynamic space-time evolution for these studies.

One option to increase thermal radiation in URHICs is a decrease of the
thermalization time, $\tau_0$, of the medium, as investigated, \eg, in
Refs.~\cite{Chatterjee:2008tp,vanHees:2011vb,Chatterjee:2012dn}.  While
the total yield generally increases above $q_T>1\, \GeV$, its slope
becomes harder and the total $v_2$ becomes smaller, both not favored by
the data. This reiterates the need for a softer radiation source with
larger $v_2$. In the following, we will stick to our default value of
$\tau_0=0.6\,\fm/c$.

\begin{figure}[!t]
\begin{minipage}{0.48\linewidth}
\includegraphics[width=\textwidth]{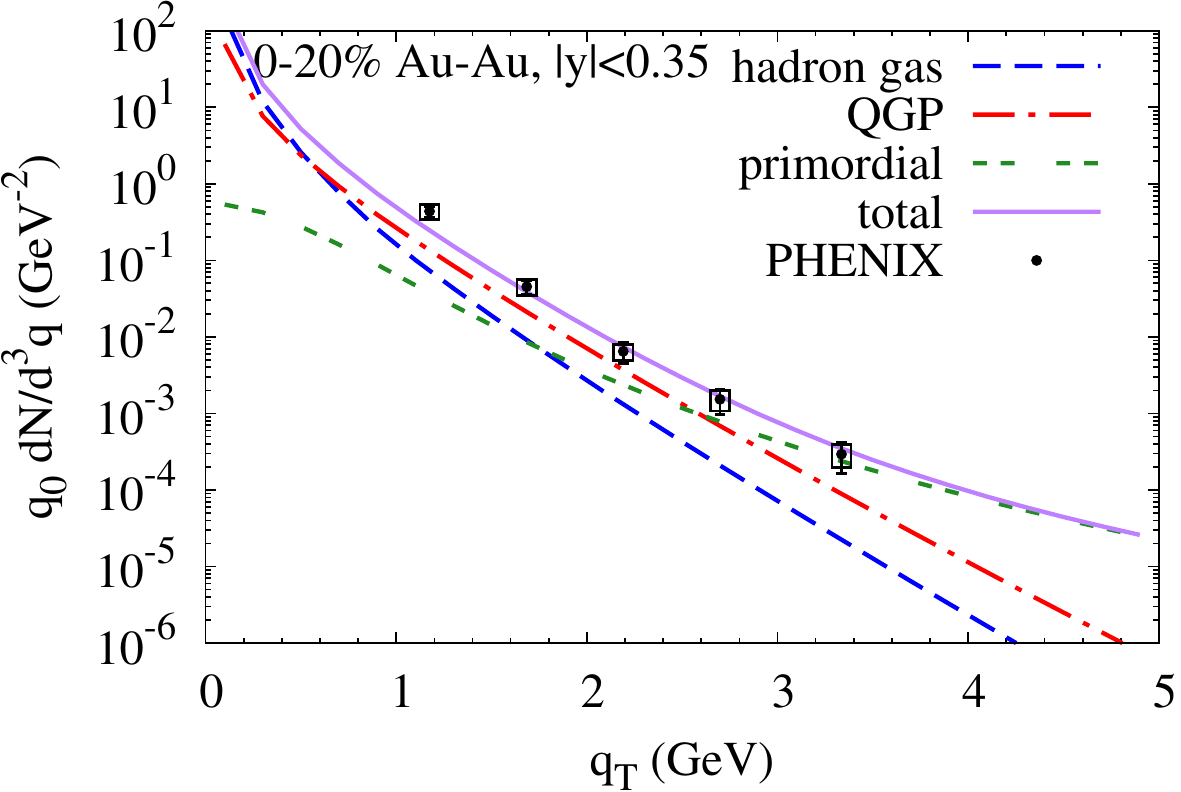}
\end{minipage}\hfill
\begin{minipage}{0.48\linewidth}
\includegraphics[width=\textwidth]{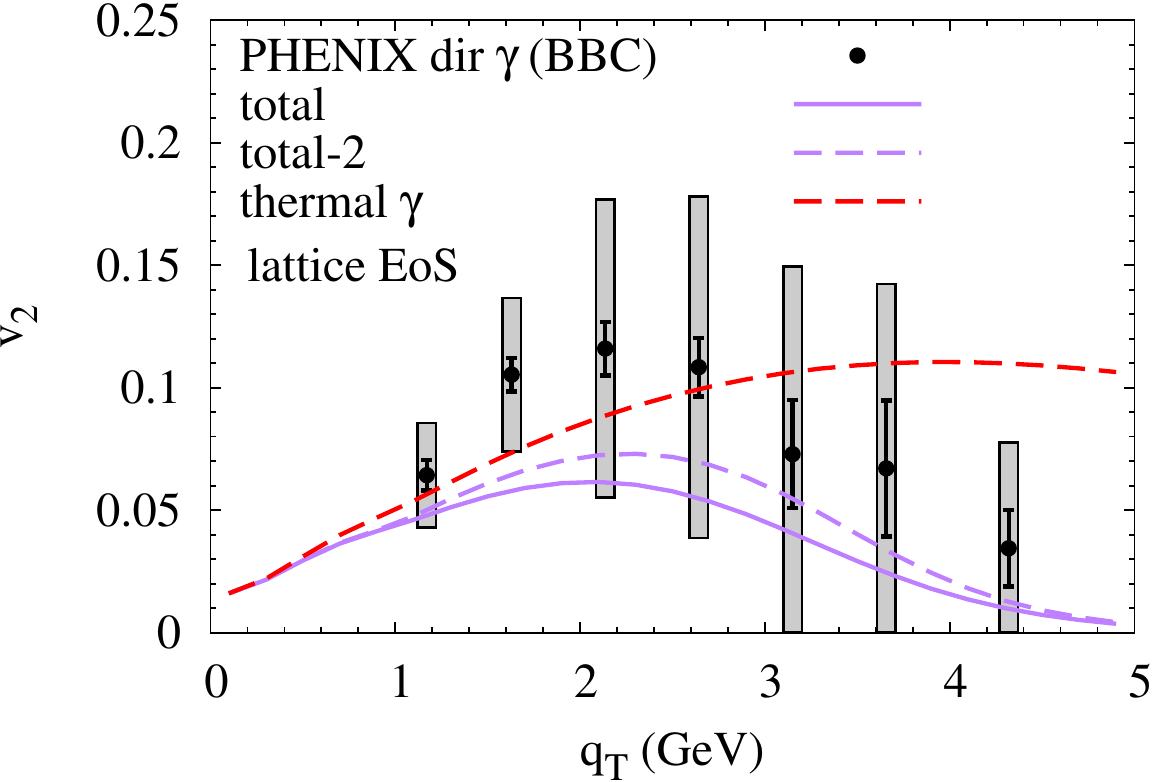}
\end{minipage}
\caption{(Color online) Direct photon spectra (left panel) and $v_2$
  (right panel) from hydrodynamics at RHIC when introducing a
  ``pseudo-critical'' enhancement of QGP and hadronic rates around
  $T_{\rm pc}$, compared to PHENIX
  data~\cite{Adare:2008fq,Adare:2011zr}; line identifications as
  Fig.~\ref{fig_fb-rhic}.}
\label{fig_rhic-ampl}
\end{figure}

\begin{figure}[!t]
\begin{minipage}{0.48\linewidth}
\includegraphics[width=\textwidth]{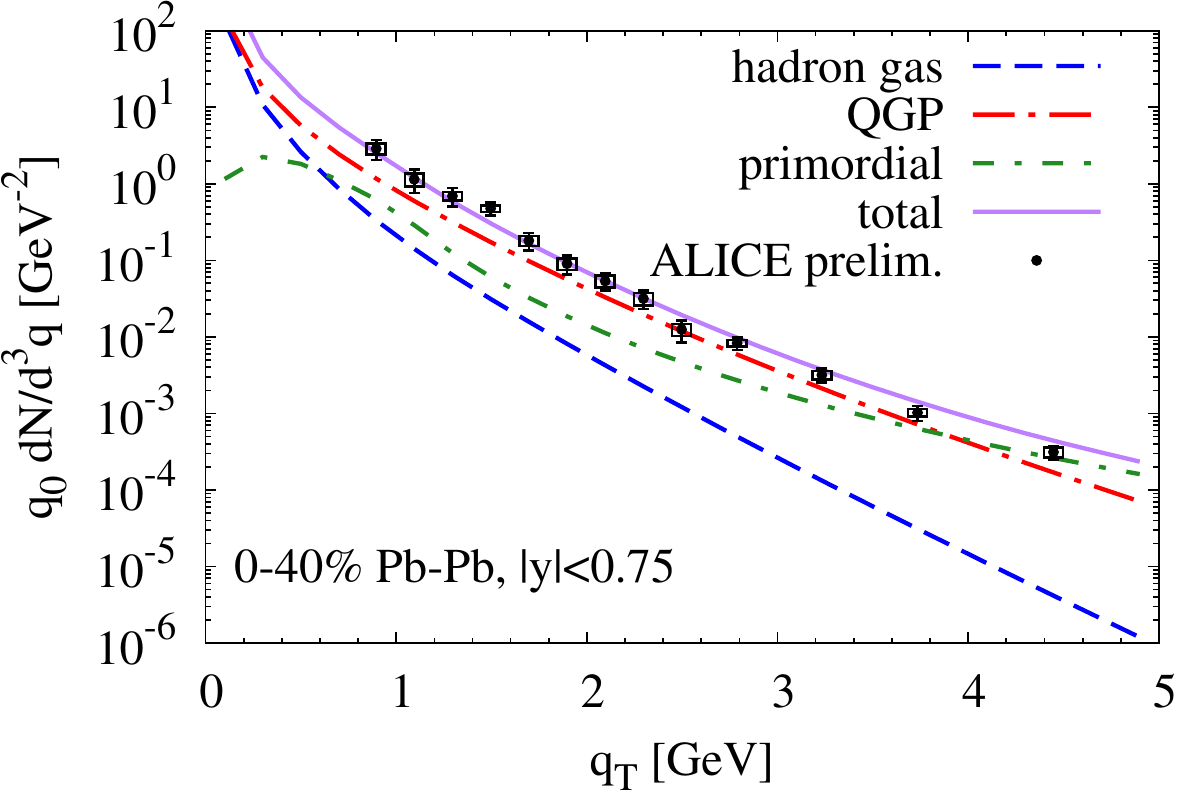}
\end{minipage}\hfill
\begin{minipage}{0.48\linewidth}
\includegraphics[width=\textwidth]{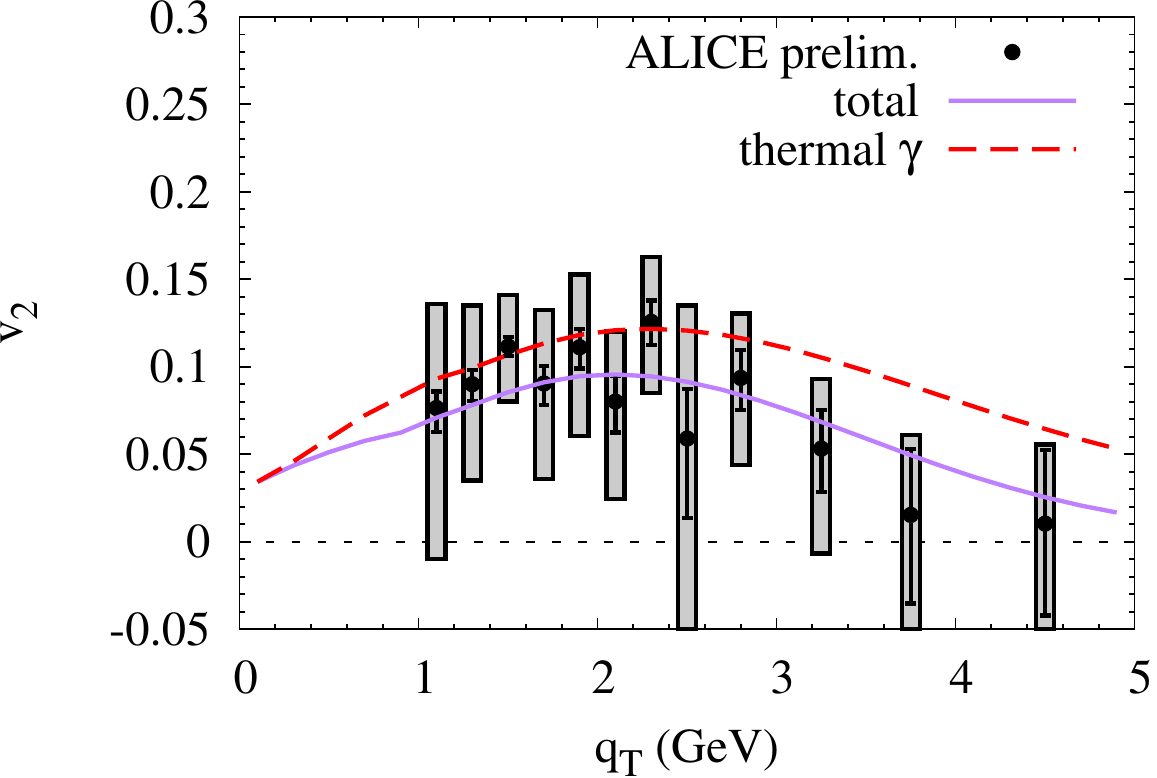}
\end{minipage}
\caption{(Color online) Direct photon spectra (left panel) and $v_2$
  (right panel) at LHC with enhanced photon rates around
  $T_\mathrm{pc}$, compared to preliminary ALICE
  data~\cite{Wilde:2012wc,Lohner:2012ct}.}
\label{fig_lhc-ampl}
\end{figure}

In Refs.~\cite{vanHees:2011vb,Rapp:2013ema,Shen:2013vja} an enhancement
of the photon emission rates in the pseudo-critical region, beyond the
default rates used above, has been conjectured. This may not be
unexpected, from a theoretical point of view. On the QGP side, the AMY
rates are based on perturbative parton rescattering, which in other
contexts tends to fall short in producing sufficient interaction
strength, \eg, in both phenomenology and lattice calculations of
$\eta/s$ or the heavy-quark diffusion
coefficient~\cite{Schafer:2009dj,Rapp:2009my}. Especially close to the
hadronization transition, confining interactions are expected to play an
important role (as, \eg, borne out of lattice calculations for the
heavy-quark free energies~\cite{Kaczmarek:2005ui}). An increase in
partonic scattering rates is a natural mechanism to also increase photon
radiation (see, \eg, Ref.~\cite{Goloviznin:2012dy}), which is quite
different from a perturbative scenario with weakly interacting
quasi-particles. On the hadronic side, an enhancement of the current
rates is conceivable as well, since the TRG rates (includiung
contributions from the in-medium $\rho$ spectral function) may not
exhaust all relevant reaction channels in hadronic resonance matter;
investigations to identify and calculate possibly important channels not
considered thus far are in progress~\cite{Holt:2014} (we note in passing
that hadronic Bremsstrahlung is an unlikely candidate since its spectrum
tends to be too soft~\cite{Liu:2007zzw}). To mimic a ``pseudo-critical"
enhancement of our default rates, we increase the latter by a baseline
factor of 2, further amplified up to a maximum factor of 3 at $T_{\rm
  pc}=170\, \MeV$, linearly ramped up from $T=140\,\MeV$ and down until
$T=200\, \MeV$ again. The results are encouraging
(cf.~Figs.~\ref{fig_rhic-ampl} and ~\ref{fig_lhc-ampl}): the description
of both PHENIX and preliminary ALICE spectra and $v_2$ improves
significantly. The calculated $v_2$ at RHIC still tends to only reach
into the lower portions of the experimental errors, but we recall that
larger hadronic contributions, as suggested by the fireball
calculations, would help to increase it further.

\begin{figure}[!t]
\begin{minipage}{0.48\linewidth}
\includegraphics[width=\textwidth]{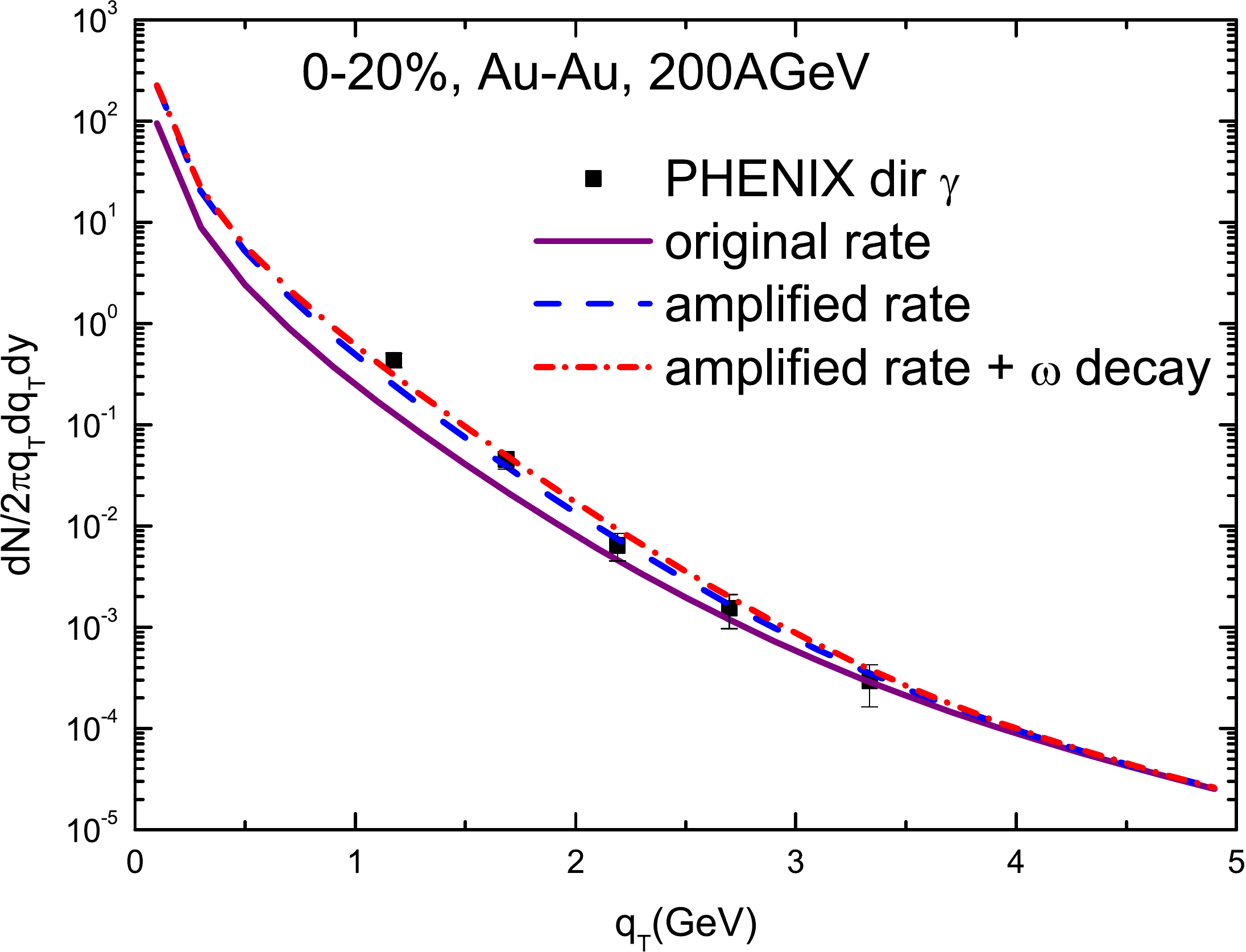}
\end{minipage}\hfill
\begin{minipage}{0.48\linewidth}
\includegraphics[width=\textwidth]{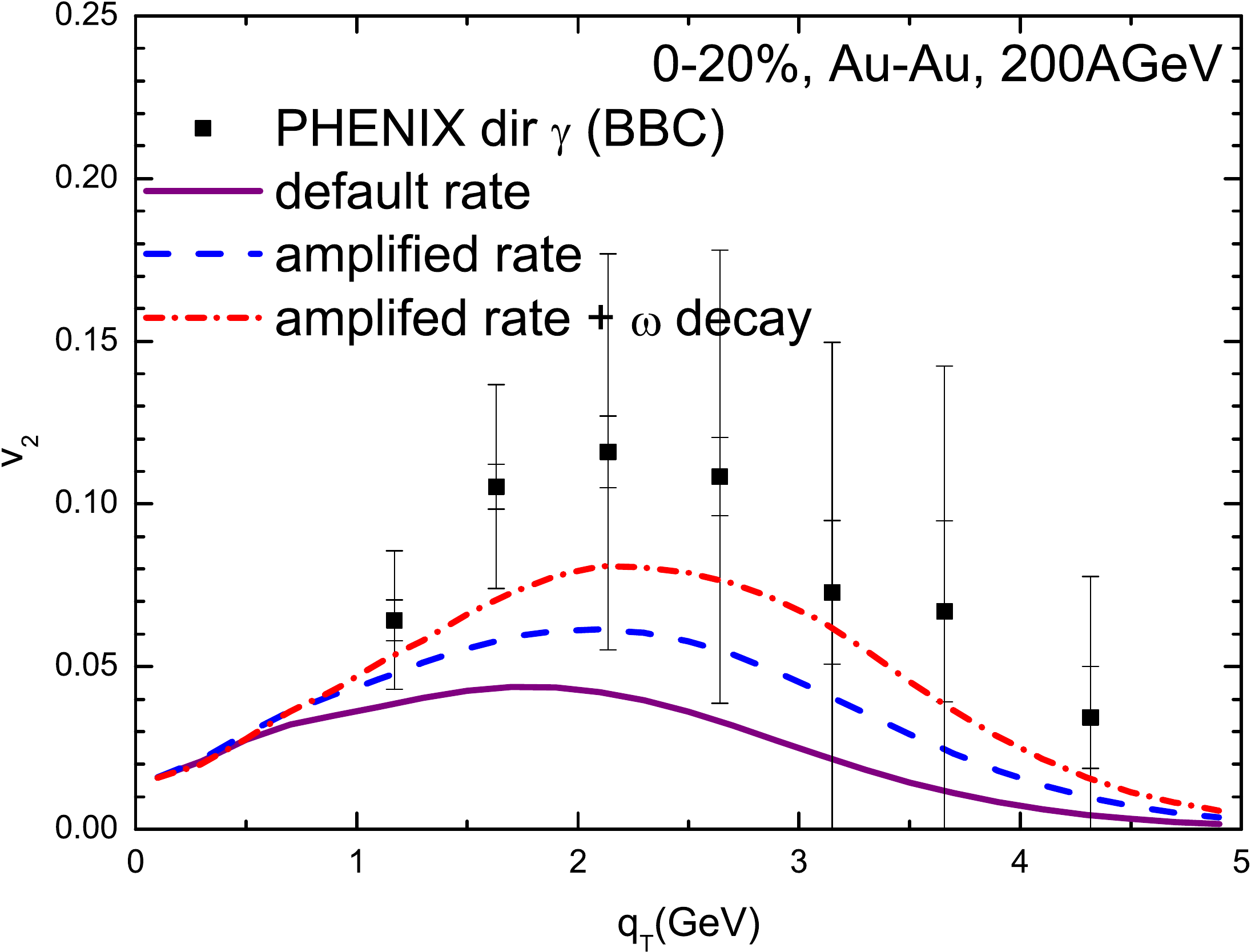}
\end{minipage}
\caption{(Color online) Direct photon spectra (left panel) and $v_2$
  (right panel) from hydrodynamics at RHIC when adding $\omega
  \rightarrow\pi^0 + \gamma$ decays at thermal freezeout to the scenario
  with enhanced rates (dash-dotted line), compared to the enhanced-rate
  (dashed line) and default-rate (solid line) scenarios, as well as
  the PHENIX data~\cite{Adare:2008fq,Adare:2011zr}; all calculations use
  the PHENIX pp baseline for the primordial component.}
\label{fig_rhic-omg}
\end{figure}

Let us briefly expand on a speculation raised in
Ref.~\cite{vanHees:2011vb}, that there might be an hitherto undetermined
uncertainty in the subtraction of the radiative $\omega \rightarrow
\pi^0\gamma$ decays, since the latter have not been explicitly measured
in the low-$q_T$ region in the Au-Au environment. For this purpose, and
to obtain an absolute upper estimate, we simply add to our thermal
spectra (calculated with the amplified rates) the photon contribution
from final-state $\omega$ decays based on our hydro $\omega$ spectra at
thermal freezeout (as a three-pion or $\rho\pi$ resonance, the $\omega$
receives a pion fugacity factor to the third power). The result of this
exercise is shown in Fig.~\ref{fig_rhic-omg}, illustrating an
appreciable effect on both spectra and $v_2$ which would still be
significant if reduced by a factor of 2.

Finally, we conduct a schematic study of the effect of quark
undersaturation in the early (Q)GP phases, as expected from gluon
saturation models~\cite{McLerran:2014hza}. Similar to earlier
calculations for thermal EM
emission~\cite{Strickland:1994rf,Srivastava:1996qd,Rapp:2000pe}, we find
the gluon Compton process, $gq\to q\gamma$, to still contribute
appreciably (and with a harder slope than in a chemically equilibrated
QGP at the same total entropy), unless the $q\bar q$ undersaturation is 
strong enough to largely suppress the early thermal yield altogether. 
This suppression would have to be made up by an even larger
enhancement in the later phases compared to what we assumed above.

\section{Summary and Conclusions}
\label{sec_sum}
In this paper we have studied the properties of thermal photon radiation
at collider energies, in an attempt to better understand recent
measurements of direct photon spectra and their elliptic flow. Using QGP
and hadronic thermal emission rates as available from the literature, we
first focused on a detailed comparison of the space-time evolution as
given by a blast-wave type fireball and ideal hydrodynamics. Both were
based on the same equation of state and fits to the same set of hadron
data using the concept of sequential freeze-out for multistrange and
light hadrons. The relevance of this concept for photon radiation lies in
a rather early saturation of $v_2$ and larger blue shifts by the time
the expanding system reaches the phase transition region (in the hydro
model, this can be realized by compact initial conditions with non-zero
radial flow). We have found that the emission characteristics of the QGP
part agree rather well between hydro and fireball, while the latter
leads to significantly larger photon radiation in the hadronic phase,
especially toward higher $q_T$. We traced this back to a slower cooling
with larger average temperatures and radial flow in the fireball, which, 
at least in part, is due to a continuous freezeout in hydrodynamics 
leading to an appreciable reduction of the ``active'' matter cells in the 
later stages of the evolution. Both evolution models clearly identify the
transition region around $T_{\rm pc}\simeq170\,\MeV$ as a key source of
thermal photon emission. After the addition of primordial photons
extrapolated from pp collisions, both hydro and fireball results tend to
be somewhat (although not dramatically) below the measured spectra and
$v_2$ in Au-Au and Pb-Pb collisions at RHIC and LHC, with a preference
for the fireball due to its larger hadronic contribution. We then
shifted our focus to the microscopic emission rates. We argued that an
enhancement of the currently employed emission rates is plausible,
especially in the pseudo-critical region where the medium is expected to
be most strongly coupled. Upon amplifying our default rates by a
baseline factor of 2, reaching up to 3 around $T_{\rm pc}\pm 30\, \MeV$,
we found that the photon results from the hydro model come rather close
to the experimental spectra and $v_2$ within current uncertainties. The
additional hadronic contributions suggested by the fireball model would
further improve the situation. Microscopic calculations of photon rates
to search for additional sources not considered thus far are underway.

\textit{ Acknowledgments.--}
We thank C.~Gale and R.J.~Fries for discussions, and gratefully
acknowledge fruitful exchanges at the EMMI RRTF workshop on
``Direct-photon flow puzzle'' organized by K.~Reygers and
J.~Stachel. This work has been supported by the U.S. National Science
Foundation through grants PHY-0969394 and PHY-1306359, by the
A.-v.-Humboldt Foundation, by NSFC grant 11305089, by the German Federal
Ministry of Education and Research (BMBF F{\"o}rderkennzeichen
05P12RFFTS), and by the Hessian initiative for excellence (LOEWE)
through the Helmholtz International Center for FAIR (HIC for FAIR).




\begin{thebibliography}{99}

\bibitem{Alam:1999sc}
  J.~Alam, S.~Sarkar, P.~Roy, T.~Hatsuda and B.~Sinha,
  Annals Phys.\  {\bf 286}, 159 (2001).

\bibitem{Peitzmann:2002}
T.~Peitzmann and M.H.~Thoma, Phys. Rep. {\bf 364}, 175 (2002).

\bibitem{Arleo:2004gn}
  F.~Arleo \etal,
  arXiv:hep-ph/0311131.

\bibitem{Rapp:2009yu}
  R.~Rapp, J.~Wambach and H.~van Hees,
  in \emph{Relativistic Heavy-Ion Physics}, edited by R.~Stock and
  Landolt B\"ornstein (Springer), New Series {\bf I/23A}, 4-1 (2010);
  [{\tt arXiv:0901.3289[hep-ph]}].

\bibitem{Gale:2009gc}
  C.~Gale,
  in \emph{Relativistic Heavy-Ion Physics}, edited by R.~Stock and
  Landolt B\"ornstein (Springer), New Series {\bf I/23A}, 6-3 (2010);
  arXiv:0904.2184 [hep-ph].

\bibitem{Chatterjee:2006}
R.~Chatterjee, E.S.~Frodermann, U.W.~Heinz and D.K.~Srivastava,
Phys. Rev. Lett. {\bf 96}, 202302 (2006).

\bibitem{Liu:2009kta}
  F.-M.~Liu, T.~Hirano, K.~Werner and Y.~Zhu,
  Phys.\ Rev.\ C {\bf 80}, 034905 (2009).

\bibitem{Holopainen:2011pd}
  H.~Holopainen, S.~R\"as\"anen and K.J.~Eskola,
  Phys.\ Rev.\ C {\bf 84}, 064903 (2011).

\bibitem{vanHees:2011vb}
  H.~van Hees, C.~Gale and R.~Rapp,
  Phys.\ Rev.\ C {\bf 84}, 054906 (2011).

\bibitem{Dion:2011pp}
  M.~Dion, J.-F.~Paquet, B.~Schenke, C.~Young, S.~Jeon and C.~Gale,
  Phys.\ Rev.\ C {\bf 84}, 064901 (2011).

\bibitem{Mohanty:2011fp}
  P.~Mohanty, V.~Roy, S.~Ghosh, S.K.~Das, B.~Mohanty, S.~Sarkar, J.~-e Alam and A.K.~Chaudhuri,
  Phys.\ Rev.\ C {\bf 85}, 031903 (2012).

\bibitem{Shen:2013cca}
  C.~Shen, U.W.~Heinz, J.-F.~Paquet, I.~Kozlov and C.~Gale,
  arXiv:1308.2111 [nucl-th].

\bibitem{Linnyk:2013wma}
  O.~Linnyk, W.~Cassing and E.~Bratkovskaya,
  Phys.\ Rev.\ C {\bf 89}, 034908 (2014).

\bibitem{Aggarwal:2000th}
M.M.~Aggarwal et~al. [WA98 Collaboration],
Phys. Rev. Lett. \textbf{85}, 3595 (2000).

\bibitem{Adare:2008fq}
  A.~Adare {\it et al.}  [PHENIX Collaboration],
  Phys.\ Rev.\ Lett.\  {\bf 104}, 132301 (2010).

\bibitem{Wilde:2012wc}
  M.~Wilde [ALICE Collaboration],
  Nucl.\ Phys.\ A {\bf 904-905}, 573c (2013).

\bibitem{Srivastava:2000pv}
  D.K.~Srivastava and B.~Sinha,
  Phys.\ Rev.\ C {\bf 64}, 034902 (2001).

\bibitem{Huovinen:2001wx}
  P.~Huovinen, P.V.~Ruuskanen and S.S.~R\"as\"anen,
  Phys.\ Lett.\ B {\bf 535}, 109 (2002).

\bibitem{Turbide:2003si}
  S.~Turbide, R.~Rapp and C.~Gale,
  Phys.\ Rev.\  C {\bf 69}, 014903 (2004).

\bibitem{Mohanty:2009cd}
  P.~Mohanty, J.K.~Nayak, J.-e Alam and S.K.~Das,
  Phys.\ Rev.\ C {\bf 82}, 034901 (2010).

\bibitem{Bauchle:2010sr}
  B.~Bauchle and M.~Bleicher,
  PoS BORMIO {\bf 2010}, 062 (2010).

\bibitem{Adare:2011zr}
  A.~Adare {\it et al.}  [PHENIX Collaboration],
  Phys.\ Rev.\ Lett.\  {\bf 109}, 122302 (2012).

\bibitem{Lohner:2012ct}
  D.~Lohner {\it et al.} [ALICE Collaboration],
  J.\ Phys.\ Conf.\ Ser.\  {\bf 446}, 012028 (2013).

\bibitem{Rapp:2000pe}
R.~Rapp, Phys. Rev. C {\bf 63}, 054907 (2001).

\bibitem{Rapp:2013nxa}
  R.~Rapp,
  Adv. High Energy Phys. {\bf 2013}, 148253 (2013).

\bibitem{Arnold:2001ms}
P.B.~Arnold, G.D.~Moore and L.G.~Yaffe,
  JHEP {\bf 0112}, 009 (2001).

\bibitem{He:2010vw}
  M.~He, R.J.~Fries and R.~Rapp,
  Phys.\ Rev.\  C {\bf 82}, 034907 (2010).

\bibitem{He:2011zx}
  M.~He, R.J.~Fries and R.~Rapp,
  Phys.\ Rev.\ C {\bf 85}, 044911 (2012).

\bibitem{Andronic:2005yp}
  A.~Andronic, P.~Braun-Munzinger and J.~Stachel,
  Nucl.\ Phys.\ A {\bf 772}, 167 (2006).

\bibitem{Stachel:2013zma}
  J.~Stachel, A.~Andronic, P.~Braun-Munzinger and K.~Redlich,
  arXiv:1311.4662 [nucl-th].

\bibitem{Rapp:2002fc}
  R.~Rapp,
  Phys.\ Rev.\ C {\bf 66}, 017901 (2002).

\bibitem{Kolb:2003dz}
  P.F.~Kolb and U.W.~Heinz,
  In R.~C.~Hwa (ed.) et al.: Quark gluon plasma, 634, [nucl-th/0305084].

\bibitem{Abelev:2007rw}
  B.I.~Abelev {\it et al.}  [STAR Collaboration],
  Phys.\ Rev.\ Lett.\  {\bf 99}, 112301 (2007).

\bibitem{Adler:2003kt}
  S.S.~Adler {\it et al.} [PHENIX Collaboration],
  Phys.\ Rev.\ Lett.\  {\bf 91}, 182301 (2003).

\bibitem{Adler:2003cb}
  S.S.~Adler {\it et al.} [PHENIX Collaboration],
  Phys.\ Rev.\ C {\bf 69}, 034909 (2004).

\bibitem{Adams:2003xp}
  J.~Adams {\it et al.}  [STAR Collaboration],
  Phys.\ Rev.\ Lett.\  {\bf 92}, 112301 (2004).

\bibitem{Pratt:2008qv}
  S.~Pratt,
  Phys.\ Rev.\ Lett.\  {\bf 102} (2009) 232301.

\bibitem{Grassi:2004dz}
  F.~Grassi,
  Braz.\ J.\ Phys.\  {\bf 35}, 52 (2005).

\bibitem{Rapp:2011is}
  R.~Rapp,
  Acta Phys.\ Polon.\ B {\bf 42}, 2823 (2011).

\bibitem{Liu:2007zzw}
  W.~Liu and R.~Rapp,
  Nucl.\ Phys.\ A {\bf 796}, 101 (2007).

\bibitem{Rapp:1999us}
R.~Rapp and J.~Wambach, Eur. Phys. J. {\bf A6}, 415 (1999).

\bibitem{vanHees:2007th}
H.~van Hees and R.~Rapp, Nucl. Phys. A \textbf{806}, 339 (2008).

\bibitem{Srivastava:2001bw}
  D.K.~Srivastava,
  Eur.\ Phys.\ J.\  C {\bf 22}, 129 (2001).

\bibitem{Shen:2013vja}
  C.~Shen, U.W.~Heinz, J.-F.~Paquet and C.~Gale,
  arXiv:1308.2440 [nucl-th].

\bibitem{Rapp:1999qu}
  R.~Rapp and C.~Gale,
  Phys.\ Rev.\ C {\bf 60}, 024903 (1999).

\bibitem{Chatterjee:2008tp}
  R.~Chatterjee and D.K.~Srivastava,
  Phys.\ Rev.\ C {\bf 79}, 021901 (2009).

\bibitem{Chatterjee:2012dn}
  R.~Chatterjee, H.~Holopainen, T.~Renk and K.J.~Eskola,
  Phys.\ Rev.\ C {\bf 85}, 064910 (2012).

\bibitem{Rapp:2013ema}
  R.~Rapp,
  PoS CPOD {\bf 2013}, 008 (2013).

\bibitem{Schafer:2009dj}
  T.~Sch\"afer and D.~Teaney,
  Rept.\ Prog.\ Phys.\  {\bf 72}, 126001 (2009).

\bibitem{Rapp:2009my}
  R.~Rapp and H.~van Hees,
  R. C. Hwa, X.-N. Wang (Ed.) Quark Gluon Plasma 4, World Scientific, 111 (2010).

\bibitem{Kaczmarek:2005ui}
  O.~Kaczmarek and F.~Zantow,
  Phys.\ Rev.\ D {\bf 71}, 114510 (2005).

\bibitem{Goloviznin:2012dy}
  V.V.~Goloviznin, A.M.~Snigirev and G.M.~Zinovjev,
  JETP Lett.\  {\bf 98} (2013) 61.

\bibitem{Holt:2014}
N.~Holt, P.~Hohler and R.~Rapp, work in progress.

\bibitem{McLerran:2014hza}
  L.~McLerran and B.~Schenke,
  arXiv:1403.7462 [hep-ph].

\bibitem{Strickland:1994rf}
  M.~Strickland,
  Phys.\ Lett.\ B {\bf 331}, 245 (1994).

\bibitem{Srivastava:1996qd}
  D.K.~Srivastava, M.G.~Mustafa and B.~M\"uller,
  Phys.\ Rev.\ C {\bf 56}, 1064 (1997).



\end{thebibliography}
\end{document}